\documentclass[10pt,conference]{IEEEtran}

\IEEEoverridecommandlockouts
\usepackage{cite}
\usepackage{amsmath,amssymb,amsfonts}
\usepackage{algorithmic}
\usepackage{graphicx}
\usepackage{textcomp}
\usepackage{xcolor}
\usepackage{enumitem}
\usepackage{makecell}
\usepackage{threeparttable}
\usepackage{multirow}
\usepackage{fontawesome}
\usepackage{subfigure}
\usepackage{verbatim}
\usepackage{balance}
\usepackage{pifont}
\usepackage{adjustbox}
\usepackage{rotating}
\usepackage[hidelinks]{hyperref}
\usepackage{booktabs}
\usepackage{inconsolata}
\usepackage{tcolorbox}
\usepackage[lined,boxed,linesnumbered,commentsnumbered,ruled]{algorithm2e}
\usepackage{color}

\newcommand{\tech}{CODA}

\newcommand{\Comment}[1]{}

\def\BibTeX{{\rm B\kern-.05em{\sc i\kern-.025em b}\kern-.08em
    T\kern-.1667em\lower.7ex\hbox{E}\kern-.125emX}}

\newcommand{\distance}{4pt}
\begin{document}

\title{Code Difference Guided Adversarial Example Generation for Deep Code Models}

\author{\IEEEauthorblockN{Zhao Tian}
\IEEEauthorblockA{\textit{College of Intelligence and} \\
\textit{Computing, Tianjin University}\\
Tianjin, China \\
tianzhao@tju.edu.cn}
\and
\IEEEauthorblockN{Junjie Chen$^\dag$}
\IEEEauthorblockA{\textit{College of Intelligence and} \\
\textit{Computing, Tianjin University}\\
Tianjin, China \\
junjiechen@tju.edu.cn}
\and
\IEEEauthorblockN{Zhi Jin}
\IEEEauthorblockA{\textit{Key Lab of High Confidence Software} \\
\textit{Technology, Peking University}\\
Beijing, China \\
zhijin@pku.edu.cn}
\thanks{$^\dag$Junjie Chen is the corresponding author.}
}

\maketitle

% 页码
% \thispagestyle{fancy}
% \renewcommand{\headrulewidth}{0pt}
% \renewcommand{\footrulewidth}{0pt} 
% \pagestyle{fancy}

\begin{abstract}
% Deep learning has been widely used to solve various code-based tasks by building deep code models based on a large number of code snippets.
% However, deep code models are still vulnerable to adversarial examples.
Adversarial examples are important to test and enhance the robustness of deep code models.
As source code is discrete and has to strictly stick to complex grammar and semantics constraints, the adversarial example generation techniques in other domains are hardly applicable.
Moreover, the adversarial example generation techniques specific to deep code models still suffer from unsatisfactory effectiveness due to the enormous ingredient search space.
In this work, we propose a novel adversarial example generation technique (i.e., \tech{}) for testing deep code models.
Its key idea is to use code differences between the target input (i.e., a given code snippet as the model input) and reference inputs (i.e., the inputs that have small code differences but different prediction results with the target input) to guide the generation of adversarial examples.
It considers both structure differences and identifier differences to preserve the original semantics.
Hence, the ingredient search space can be largely reduced as the one constituted by the two kinds of code differences, and thus the testing process can be improved by designing and guiding corresponding equivalent structure transformations and identifier renaming transformations.
Our experiments on 15 deep code models demonstrate the effectiveness and efficiency of \tech{}, the naturalness of its generated examples, and its capability of enhancing model robustness after adversarial fine-tuning.
For example, \tech{} reveals 88.05\% and 72.51\% more faults in models than the state-of-the-art techniques (i.e., CARROT and ALERT) on average, respectively.
% improves the state-of-the-art techniques (i.e., CARROT and ALERT) by 88.05\% and 72.51\% on average in terms of the number of revealed inconsistencies, respectively.
\end{abstract}

% \begin{IEEEkeywords}
% xxxx, xxxx, xxxx
% \end{IEEEkeywords}

% \vspace{-1mm}
\section{Introduction}
\label{sec:intro}

% DL is widely-used in SE, involving many SE tasks
In recent years, deep learning (DL) has been widely used to solve code-based software engineering tasks, such as code clone detection~\cite{white2016deep} and vulnerability prediction~\cite{zhou2019devign}
% , and code completion~\cite{li2018code}, 
by building DL models based on a large amount of training code snippets (called \textit{deep code models}).
Indeed, deep code models have achieved notable performance and largely promoted the process of software development and maintenance~\cite{alon2018code2seq,alon2019code2vec,feng2020codebert,guo2021graphcodebert}. 
% In particular, some industrial products on deep code models have been released and received extensive attention, such as AlphaCode~\cite{li2022competition} and Codex~\cite{chen2021evaluating}.
In particular, some industrial products on deep code models have been released and received extensive attention, such as AlphaCode~\cite{li2022competition} and Copilot~\cite{copilot2023}.

% Recently, deep learning (DL) has achieved state-of-the-art performance on many source code processing tasks, such as vulnerability prediction, functionality classification, code clone detection, etc. 
% Some of these technologies can improve the efficiency of software and system development, and provide solutions for the industry.

% Like DL in other fields, the robustness of deep code models should receive attention. adversarial attack is widely-used, but existing techniques in other fields cannot be applicable to code models.

Like DL models in other areas (e.g., image processing)~\cite{kurakin2018adversarial,wang2021prioritizing,shen2022natural,yan2023revisiting}, the robustness of deep code models is critical~\cite{zhang2022towards,li2022cctest}.
% In this area, adversarial examples are important to evaluate and increase the model robustness~\cite{zhang2020generating,yang2022natural}.
As demonstrated by the existing work~\cite{zhang2020generating,yang2022natural}, adversarial examples are important to test and enhance the model robustness.
Specifically, adversarial examples can test a deep code model to reveal faults in it by comparing the prediction results on adversarial examples and that on the original input generating these adversarial examples.
Such adversarial examples are called \textit{fault-revealing examples} for ease of presentation, which can be used to augment training data for further enhancing the model robustness.
Therefore, improving test effectiveness through generating fault-revealing examples is very important.

% However, the existing adversarial attack techniques in other areas are hardly applicable to deep code models.
However, the existing adversarial example generation techniques in other areas are hardly applicable to deep code models.
This is because they tend to perturb an input in continuous space, while the inputs (i.e., source code) for deep code models are discrete.
Moreover, source code has to strictly stick to complex grammar and semantics constraints, i.e., the adversarial example generated from an original input should have no grammar errors and preserve the original semantics.

% The performance in terms of accuracy and robustness of the DL source code processing model is critical.
% Although DL models have been shown to achieve competitive performance in terms of accuracy, the question of robustness remains unresolved. 
% Adversarial examples can be obtained by adding tiny perturbations to the original code that are crafted by the adversary and imperceptible to humans. 
% Such adversarial examples that do not change the semantics of the original code can lead to DL model prediction errors. 
% If abused by attackers, it may also bring potential security risks to downstream tasks, which makes the robustness of the model even more important. 
% This security issue can be fatal to the entire system.

% There are many techniques for generating adversarial examples in the computer vision (CV) field, and their basic idea is to add specially crafted noise to continuous input objects to alter the model's predictions.
% Unlike images, the source code space is discrete and must strictly follow strict lexical and syntactic constraints. 

% There are some adversarial attack techniques specific to code models, but suffer from limitations.
% Indeed, some adversarial attack techniques specific to deep code models have been proposed recently, such as MHM~\cite{zhang2020generating}, CARROT~\cite{zhang2022towards}, and ALERT~\cite{yang2022natural}.
Indeed, some adversarial example generation techniques specific to deep code models have been proposed recently, such as MHM~\cite{zhang2020generating}, CARROT~\cite{zhang2022towards}, and ALERT~\cite{yang2022natural}.
In general, they share two main steps:
(1) designing semantic-preserving code transformation rules,
and (2) searching ingredients from the space defined by the rules (i.e., ingredients are the elements required by transformation rules) 
for transforming an original input (called \textit{target input}) to a semantic-preserving adversarial example.
For example, CARROT designs two semantic-preserving code transformation rules (i.e., identifier renaming and dead code insertion), and uses the hill-climbing algorithm to search for the ingredients from the entire space with the guidance of gradients and model prediction changes.
ALERT considers the rule of identifier renaming, and uses the naturalness and model prediction changes to guide the ingredient search process.

Although some of them have been demonstrated effective to some degree, they still suffer from major limitations:
\begin{itemize}
\item The ingredient space defined by code transformation rules is enormous.
For example, all valid identifier names could be the ingredients for identifier renaming transformation.
% Hence, searching for the ingredients that can help attack the target model successfully is challenging.
Hence, searching for the ingredients that facilitate generating fault-revealing examples is challenging.
% The existing techniques tend to utilize the changes of model prediction results after performing semantic-preserving transformations on the target input greedily to guide the search process, which is likely to fall into local optimum in the enormous space and thus limits their attack effectiveness~\cite{yang2022natural}.
The existing techniques tend to utilize model prediction changes after performing transformations on the target input greedily to guide the search process, which is likely to fall into local optimum and thus limits test effectiveness~\cite{yang2022natural}.
% zhang2020generating,yang2022natural,zhang2022towards

% \item Frequently invoking the target model could affect the efficiency of adversarial attack techniques to some degree~\cite{zhang2022towards,yang2022natural}, as model invocation is the most costly part during the attack process.
\item Frequently invoking the target model could affect test efficiency via adversarial example generation~\cite{zhang2022towards,yang2022natural}, as model invocation is the most costly part during testing.
% Moreover, when the model is deployed remotely, frequent model invocations could be identified as malicious attacks and thus lead to blocking access to the model~\cite{yang2022natural}.
However, the existing techniques often involve frequent model invocations due to calculating gradients or guiding the search direction via model prediction. 
% \tz{A Reviewer said this is not important.}\jj{does any paper say it is an important point?}\tz{Both ALERT and CARROT think this is a very important point.}

\item Developers care about natural semantics of code since it can assist human comprehension on detected faults in models~\cite{casalnuovo2020theory}.
Hence, ensuring the naturalness of generated adversarial examples is important.
However, all the existing techniques (except ALERT~\cite{yang2022natural}) do not consider this factor.
For example, CARROT designs the rule of dead code insertion, but it may largely damage the naturalness of generated examples (especially when a large amount of dead code is inserted). 
% \tz{reviewer 1: for data argumentation technique, naturalness is of lesser importance}
\end{itemize}

% Overall, a more effective adversarial attack technique specific to deep code models should enhance the effectiveness by improving the ingredient search process, and guarantee the naturalness of generated adversarial examples as much as possible and the times of model invocations as few as possible.
Overall, a more effective adversarial example generation technique for testing deep code models should improve the ingredient search process, and guarantee the times of model invocations as few as possible and the naturalness of generated adversarial examples as much as possible.
Our work proposes such a technique, called \textbf{\tech{}} (\textbf{CO}de \textbf{D}ifference guided \textbf{A}dversarial example generation).
% \tz{CODT sounds not good?}

% \jin{\begin{CJK}{UTF8}{gkai}
% 读起来感觉这两段需要再精炼一点，重点把xx说清楚，说到位。
% 1. 攻击的有效性  2.攻击的隐蔽性（即自然性），
% 这两点宜再明确一些
% \end{CJK}}

% \jin{1. this paragraph}
% To improve the attack effectiveness, the \textit{key idea} of \tech{} is to take \textit{reference inputs} (that have small code differences with the target input but have different prediction results) as the guidance for generating adversarial examples.
To improve test effectiveness, the key idea of \tech{} is to take \textit{reference inputs} (that have small code differences with the target input but have different prediction results) as the guidance for generating adversarial examples.
This is an innovative perspective and closely utilizes the unique characteristics of deep code models.
% \textit{\underline{Why does this idea work?}}
With reference inputs, the ingredient space can be largely reduced.
Specifically, reference inputs can be regarded as \textit{invalid} fault-revealing adversarial examples generated from the target input, where ``invalid'' refers to altering the original semantics and ``fault-revealing'' refers to producing different prediction results.
% That is, the code differences brought by reference inputs over the target input contribute to the invalid but \textit{successful} attack to a large extent.
That is, the code differences brought by reference inputs over the target input contribute to the generation of invalid but \textit{fault-revealing} example to a large extent.
Hence, if we extract the ingredients from the code differences to support semantic-preserving transformations on the target input, their code differences can be gradually reduced without altering the original semantics, and thus a \textit{valid} fault-revealing example is likely to be generated.
In this way, the ingredient space is reduced as the one constituted by only code differences between reference inputs and the target input, and thus the search process can be improved.

Based on the key idea,
% To preserve the semantics of the target input during the attack process, 
\tech{} considers code structure differences and identifier differences to support ingredient extraction for equivalent structure transformations and identifier renaming transformations.
% It can preserve the original semantics during the attack process.
It can preserve the original semantics during the generation process.
% Although some of transformation rules have been proposed before, the code-difference-guided transformation process designed in \tech{} is novel and can significantly improve the attacking effectiveness of existing transformation rules.
Equivalent structure transformations (e.g., transforming a {\tt for} loop to an equivalent {\tt while} loop) do not affect the naturalness of generated examples, and thus \tech{} first applies this kind of transformations to reduce code differences for generating adversarial examples.
% Then, identifier renaming transformations are applied to further reduce code differences to improve the attack effectiveness.
Then, identifier renaming transformations are applied to further reduce code differences to improve test effectiveness.
To ensure the naturalness of generated examples by this kind of transformations, 
% and minimize the perturbations on the target input, 
\tech{} measures semantic similarity between identifiers
% the semantic similarity between each identifier of the target input and each ingredient identifier 
for guiding iterative transformations.
% Note that we do not emphasize the novelty in these transformation rules since some of them have been proposed before, and the main novelty lies in the code-difference-guided transformation process in \tech{}, which is the key to improve the test effectiveness of these transformations.
Note that we do not emphasize the novelty in these transformation rules since some of them have been proposed before, and the main novelty lies in the code-difference-guided transformation process in \tech{}, which is the key to improve the test effectiveness with these transformations.
In particular, \tech{} just involves necessary model invocations to check whether the generated example reveals a fault, without extra gradient calculation and a large amount of model prediction for guiding the search process.

We conducted an extensive study to evaluate \tech{} based on three popular pre-trained models (i.e., CodeBERT~\cite{feng2020codebert}, GraphCodeBERT~\cite{guo2021graphcodebert}, and CodeT5~\cite{wang2021codet5}) and five code-based tasks. 
% (i.e., vulnerability prediction~\cite{zhou2019devign}, clone detection~\cite{wei2017supervised}, authorship attribution~\cite{alsulami2017source}, functionality classification~\cite{zhang2019novel}, and defect prediction~\cite{zhang2022towards}).
In total, we used 15 subjects.
Our results demonstrate the effectiveness and efficiency of \tech{}.
% For example, on average across all the subjects, \tech{} improves the two state-of-the-art adversarial attack techniques specific to deep code models (i.e., CARROT~\cite{zhang2022towards} and ALERT~\cite{yang2022natural}) by 88.05\% and 72.51\% in terms of the attack success rate, respectively.
For example, on average across all the subjects, \tech{} revealed 88.05\% and 72.51\% more faults in models than the two state-of-the-art adversarial example generation techniques (i.e., CARROT~\cite{zhang2022towards} and ALERT~\cite{yang2022natural}), respectively.
% improves the two state-of-the-art adversarial example generation techniques (i.e., CARROT~\cite{zhang2022towards} and ALERT~\cite{yang2022natural}) by 88.05\% and 72.51\% in terms of the number of revealed inconsistencies, respectively.
% The time spent by \tech{} on completing the attack process for the 15 subjects is 196.96 hours, while those by CARROT and ALERT are 290.87 hours and 374.51 hours, respectively.
The time spent by \tech{} on completing the testing process for the 15 subjects is 196.96 hours, while those by CARROT and ALERT are 290.87 hours and 374.51 hours, respectively.
Furthermore, we investigated the value of the generated adversarial examples by using them to enhance the robustness of the target model via an adversarial fine-tuning strategy.
The results show that the models after fine-tuning with the examples generated by \tech{} can reduce 62.19\%, 65.67\%, and 73.95\% of faults revealed by CARROT, ALERT, and \tech{} on average, respectively.

To sum up, our work makes four major contributions:
\begin{itemize}
    % \item \textbf{Novel Perspective}. We propose a novel perspective of utilizing code differences between reference inputs and the target input to guide the adversarial attack process for deep code models.
    \item \textbf{Novel Perspective}. We propose a novel perspective of utilizing code differences between reference inputs and the target input to guide the fault-revealing example generation process for testing deep code models.

    \item \textbf{Tool Implementation}. We implement \tech{} following the novel perspective by (1) measuring code structure and identifier differences and (2) designing and guiding corresponding semantic-preserving code transformations.

    \item \textbf{Performance Evaluation}. We conducted an extensive study on three popular pre-trained models and five code-based tasks, demonstrating the effectiveness and efficiency of \tech{} over two state-of-the-art techniques.

    \item \textbf{Public Artifact}. We released all the experimental data and our source code at the project homepage~\cite{coda2022} for experiment replication, future research, and practical use.
    
    % We propose a new attack technique (\tech{}) that generates adversarial examples by changing the code structure and variable names to eliminate differences between inputs with different labels. 
    % This can greatly reduce the search space and improve the attack success rate.
    % According to our knowledge, we are the first adversarial attack technique that considers code structure and the relationship of inputs with different labels.
    
    % \item \tech{} require less invocation of DL model during the attack process, which is less time-consuming and practical.
    
    % \item \tech{} devises equivalent transformations of code structure, which can generate natural adversarial examples, and achieve higher attack success rate and efficiency.
    
    % \item We develop and publish our tools and data to facilitate future research and practical use. Please find them at:  \textbf{\url{https://}}.\tz{to be updated}
\end{itemize}

\begin{figure*}[t!]
    \centering
    \includegraphics[width=1.0\linewidth]{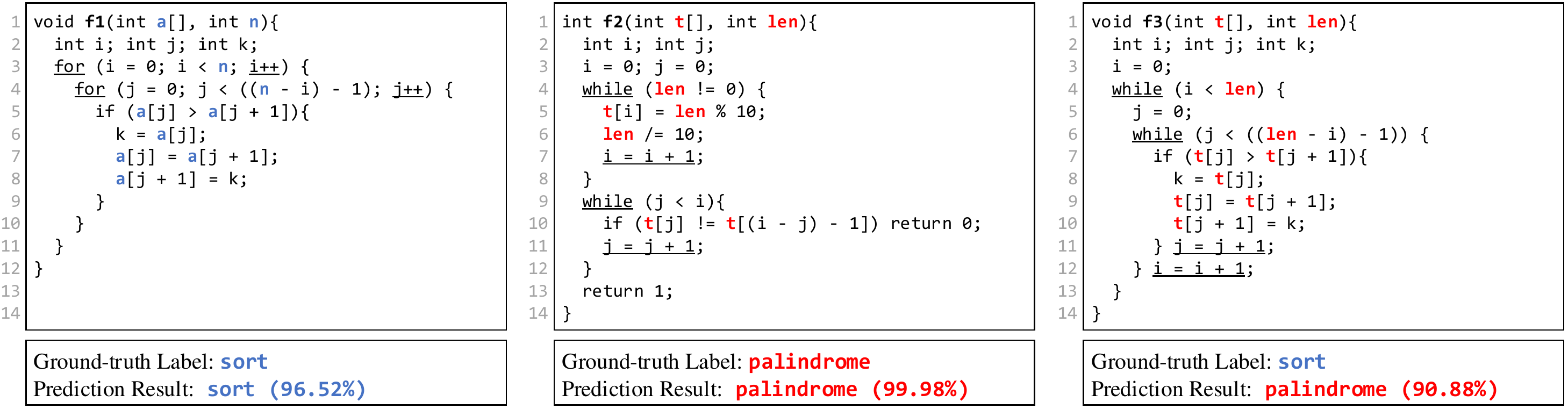}
    % \vspace{-6.5mm}
    % \vspace{-3mm}
    \caption{An illustrating example (the target input \textit{f1}, a reference input \textit{f2}, and a fault-revealing adversarial example \textit{f3} generated from \textit{f1})}
    \label{fig:example}
    
\end{figure*}

% \vspace{-1mm}
\section{Background and Motivation}
\label{sec:background}
% \jj{introduce deep code models and define our problem.}
% In this section, we first introduce the background of deep code models (Section~\ref{sec:deepcodemodel}), define our problem (Section~\ref{sec:problem}), and motivate our key idea with an example (Section~\ref{sec:motivation}).

% In this section, we introduce deep code models and give a formal definition of the source code classification. 
% Then, we introduce the procedure of adversarial attack. 

% \vspace{-1mm}
\subsection{Deep Code Models}
\label{sec:deepcodemodel}
% In the area of software engineering, 
DL has been widely used to process source code~\cite{wei2017supervised,li2018code,zhou2019devign,huang2018api,tian2022learning,gao2023vectorizing,kang2021apirecx}.
% In particular, 
Some pre-trained DL models have been built based on a large number of code snippets, among which CodeBERT~\cite{feng2020codebert}, GraphCodeBERT~\cite{guo2021graphcodebert}, and CodeT5~\cite{wang2021codet5} are three state-of-the-art pre-trained models.
% CodeBERT learns features from bimodal data in the form of programming languages and natural languages, GraphCodeBERT takes into consideration the code structure and data flow information, and 
% CodeT5 is a unified encoder-decoder transformer-based model that learns code semantics via an identifier-aware pre-training objective.
% \tz{or "CodeT5 considers the token type information in code."?}
% Same as the existing work~\cite{yang2022natural}, we used CodeBERT and GraphCodeBERT in our evaluation, and also used the more recent CodeT5 (Section~\ref{sec:design}).
The pre-trained models have brought breakthrough changes to many code-based tasks~\cite{karmakar2021pre}, including both classification and generation tasks, by fine-tuning them on the datasets of the corresponding tasks.
The former makes classification based on the given code snippets (e.g., vulnerability prediction~\cite{zhou2019devign}), while the latter produces a sequence of information based on code snippets or natural language descriptions (e.g., code completion~\cite{li2018code}).
% Following most of the existing work on attacking deep code models~\cite{zhang2020generating,zhang2022towards,yang2022natural}, our work also focuses on the classification tasks and takes the generation tasks as our future work.
Following most of the existing work on generating adversarial examples for deep code models~\cite{zhang2020generating,zhang2022towards,yang2022natural}, our work also focuses on the classification tasks and takes the generation tasks (targeted by ACCENT~\cite{zhou2022adversarial}, CCTest~\cite{li2022cctest}, etc.) as our future work.
% In our study, we adopted all the tasks used in the studies of evaluating the state-of-the-art attack techniques (i.e., CARROT~\cite{zhang2022towards} and ALERT~\cite{yang2022natural}), i.e., five classification tasks including vulnerability prediction, clone detection, authorship attribution, functionality classification, and defect prediction.
In our study, we adopted all the five classification tasks used in the studies of evaluating the state-of-the-art adversarial example generation techniques (i.e., CARROT~\cite{zhang2022towards} and ALERT~\cite{yang2022natural}).
\subsection{Problem Definition}
\label{sec:problem}
Given a code snippet $x$ that is processed as the required format by the target deep code model $\mathcal{M}$ (e.g., abstract syntax trees required by code2seq~\cite{alon2018code2seq}, control-flow graphs required by DGCNN~\cite{phan2017convolutional}, or data-flow graphs required by GraphCodeBERT~\cite{guo2021graphcodebert}), $\mathcal{M}$ can predict a probability vector for $x$, each element in which represents the probability classifying $x$ to the corresponding class.
The class with the largest probability is the final prediction result of $\mathcal{M}$ for $x$.
If the prediction result is different from the ground-truth label (denoted as $y$) of $x$, it means that $\mathcal{M}$ makes a wrong prediction on $x$; otherwise, $\mathcal{M}$ makes a correct prediction.

Same as the existing work~\cite{zhang2022towards,yang2022natural}, 
our goal is to improve test effectiveness through more effectively generating fault-revealing examples, which subsequently can be used to enhance model robustness.
% our goal is to generate an inconsistency-revealing adversarial example for each target input so as to test and enhance the model robustness.} \tz{Reviewers may ask why don't you generate more adversarial examples to reveal more inconsistencies?}
% As source code is discrete and has to stick to grammar and semantics constraints, the existing adversarial example generation techniques proposed in other domains are hardly applicable.
% \yl{New paragraph, together with the next one as formal definition}
% The existing attack techniques specific to deep code models always generate adversarial examples from a target input by performing a series of semantic-preserving code transformations~\cite{zhang2020generating,zhang2022towards,yang2022natural}, which is also followed by our work.
The existing techniques specific to deep code models always generate adversarial examples from a target input by performing a series of semantic-preserving code transformations~\cite{zhang2020generating,zhang2022towards,yang2022natural}, which is also followed by our work.
% Our work also follows the same workflow\tz{same workflow?}.
For ease of understanding, we formally present our problem as finding $x'$ ($x' \in \epsilon \wedge y = \mathcal{M}(x) \neq \mathcal{M}(x')$)
% $\{x'|x' \in \epsilon \wedge y = \mathcal{M}(x) \neq \mathcal{M}(x')\}$ 
from a target input $x$ for the target model $\mathcal{M}$.
Here, $\epsilon$ is the universal set of code snippets that satisfy the grammar constraints and preserve the semantics of $x$.
$y = \mathcal{M}(x)$ means that we just regard the test inputs on which $\mathcal{M}$ makes correct predictions as target inputs since analyzing robustness upon such inputs is more meaningful following the existing work~\cite{yang2022natural,zhang2022towards}, where $\mathcal{M}(x)$ refers to the prediction result of $\mathcal{M}$ on $x$.
$\mathcal{M}(x) \neq \mathcal{M}(x')$ means that $x'$ reveals a fault in $\mathcal{M}$, that is, it is a fault-revealing example generated from $x$.
Besides, an effective adversarial example generation technique should be \textit{efficient} to find $x'$ and ensure the \textit{naturalness} of $x'$ (i.e., natural to human comprehension~\cite{yang2022natural}), which are indeed carefully considered by our proposed technique.

\subsection{Motivating Example}
\label{sec:motivation}
% \tz{Fig. \ref{fig:example}: a real-world example from Functionality Classification task, ALERT and CARROT cannot attack \texttt{f1} successfully.}
% ========================

We use a real-world example (simplified for ease of illustration) to motivate our key idea: utilizing code differences between reference inputs and the target input to guide the generation of adversarial examples.
In Figure~\ref{fig:example}, the first code snippet \textit{f1} is the target input from the test set of the functionality classification task~\cite{zhang2019novel}, and the two state-of-the-art techniques (i.e., CARROT~\cite{zhang2022towards} and ALERT~\cite{yang2022natural}) do not generate fault-revealing examples from it for the deep code model CodeBERT since they can fall into local optimum in the enormous ingredient space.
The second code snippet \textit{f2} is a reference input from the training set of this task, which has the different label
% \yl{since its in training data, use different label} 
with \textit{f1}.

As presented before, \textit{f2} can be regarded as an \textit{invalid} fault-revealing example from \textit{f1}, as they are semantically inconsistent but have different prediction results.
% \yl{Actually, the model successfully predict f2 as palindrome, so for the palindrome class, it is a unsuccessful attack if we regard this as an adversarial example.}
The code differences between \textit{f1} and \textit{f2} mainly contribute to this phenomenon.
From this perspective, to generate a \textit{valid} fault-revealing example (denoted as \textit{f3}) from \textit{f1}, we should perform semantic-preserving code transformations on \textit{f1}, and the transformations should reduce the code differences between \textit{f1} and \textit{f2} to alter the prediction result of the model from \textit{f1}.
That is, the ingredients supporting these transformations should be extracted from the code differences brought by \textit{f2}.
With the guidance of code differences, by performing equivalent structure transformations on \textit{f1} (i.e., transforming {\tt for} loops to {\tt while} loops, where {\tt while} loops are the used loop structure in \textit{f2}) and identifier remaining transformations (i.e., renaming {\tt a} and {\tt n} to {\tt t} and {\tt len} respectively, where {\tt t} and {\tt len} are the used identifier names in \textit{f2}), 
\textit{f3} is generated as shown in the third code snippet in Figure~\ref{fig:example} and indeed reveals a fault in the model, i.e., making a wrong prediction ({\tt palindrome}) with a high confidence (90.88\%).

Based on the code differences between the target input and the reference input, the ingredient space is largely reduced.
For example, the ingredient space defined by identifier renaming transformations is reduced from all valid identifier names (i.e., almost infinite) to the identifier names occurring in the reference input but not in the target input (i.e., only two identifiers in this example).
% Hence, it helps improve the ingredient search process and thus improves the attack effectiveness.
Hence, it helps improve the ingredient search process and thus improves test effectiveness.
% On the other hand, too small ingredient space may lose too many ingredients useful to successful attacks, and thus we will select a set of reference inputs (rather than only one reference input) for guiding the attack process so as to balance the ingredient-space size and the number of useful ingredients.
On the other hand, too small ingredient space may lose too many ingredients useful to generate fault-revealing examples, and thus we will select a set of reference inputs (rather than only one reference input) for guiding the generation process so as to balance the ingredient-space size and the number of useful ingredients.

\section{Approach}
\label{sec:approach}
\begin{figure}[t]
    \centering
    \includegraphics[width=1.0\linewidth]{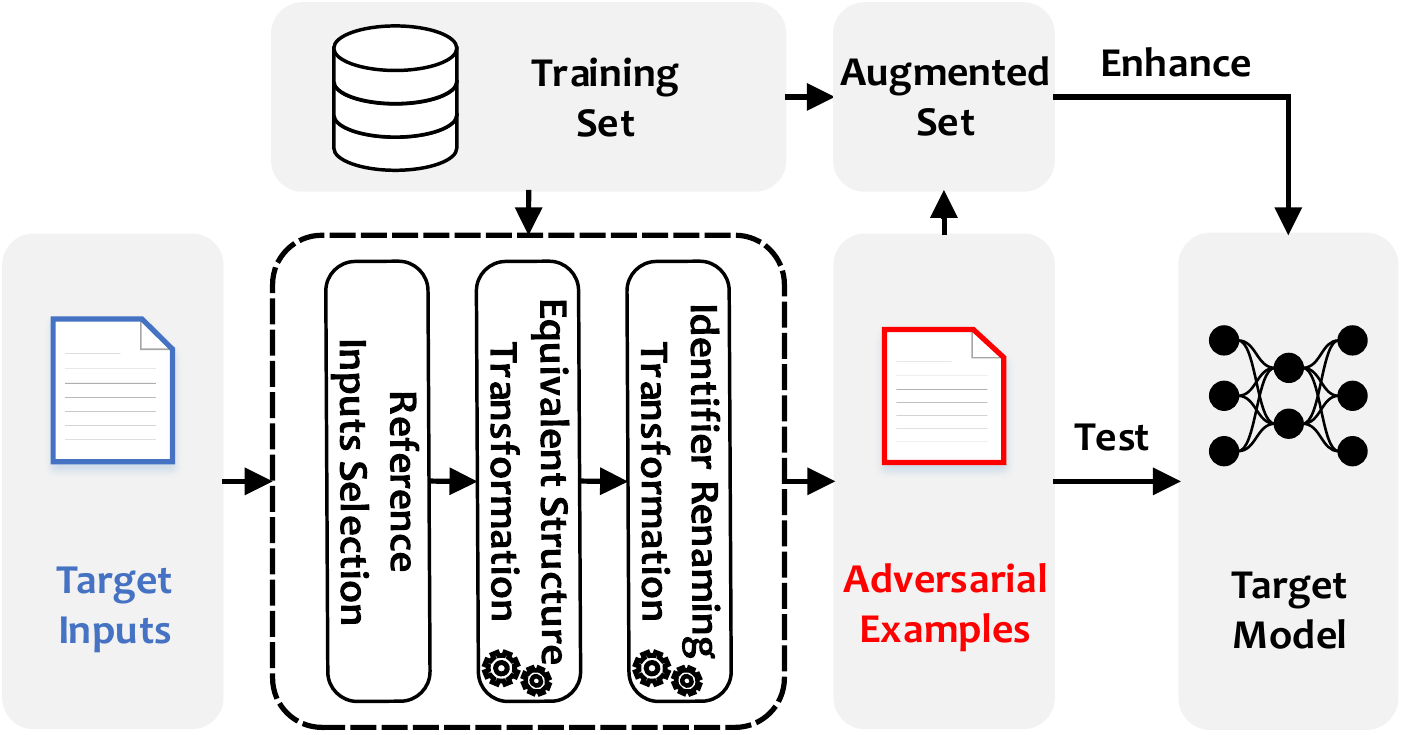}
    % \vspace{-3mm}
    \caption{Overview of \tech{}.}
    \label{fig:overview}
    % \vspace{-5mm}
\end{figure}

% \vspace{-0.5mm}
\subsection{Overview}
\label{sec:overview}
% \jj{could be reduced.}
% In this work, we propose a novel perspective to attack deep code models more effectively, which utilizes code differences between reference inputs and the target input to guide the generation of adversarial examples.
We propose a novel perspective to generate adversarial examples for better testing deep code models, which utilizes code differences between reference inputs and the target input to guide the generation process.
% test deep code models more effectively, which utilizes code differences between reference inputs and the target input to guide the generation of adversarial examples.
% From this perspective, we design an effective attack technique, called \tech{}.
From this perspective, we design an effective adversarial example generation technique, called \tech{}.
% Specifically, the code differences brought by reference inputs provide effective ingredients for altering the prediction result of the target input by transforming it with these ingredients, which can contribute to successful attacks in \tech{}.
Specifically, the code differences brought by reference inputs provide effective ingredients for altering the prediction result of the target input by transforming it with these ingredients, which can contribute to the generation of fault-revealing examples in \tech{}.
However, as the semantics of reference inputs and the target input are different, the ingredients from some kinds of code differences can alter the original semantics, which is not allowed by the adversarial examples for deep code models.
% Hence, in \tech{}, we consider structure differences and identifier differences for measuring code differences between them, which can preserve the original semantics during the attack process.
Hence, in \tech{}, we consider structure differences and identifier differences for measuring code differences between them, which can preserve the original semantics in generation.
In this way, the ingredient space can be reduced as the one constituted by the two kinds of code differences, and thus the ingredient search process (for generating adversarial examples) can be largely improved.

% In fact, not all the inputs that have different prediction results with the target one, can be regarded as effective reference inputs for improving the adversarial attack process.
% \tz{We may delete some sentences, and the following subsections have repeated sentences} 

In fact, not all inputs that have different prediction results with the target one, can be regarded as effective reference inputs for improving test effectiveness.
In other words, different inputs could have different degrees of capabilities for reducing the ingredient search space and providing effective ingredients for altering the prediction result of the target input.
% Hence, the first step in \tech{} is to select effective reference inputs for the target input in order to improve the attack effectiveness (Section~\ref{sec:ReferenceInputsSelection}).
Hence, the first step in \tech{} is to select effective reference inputs for the target input in order to improve test effectiveness (Section~\ref{sec:ReferenceInputsSelection}).
Based on the selected reference inputs, \tech{} then measures structure differences and identifier differences over the target input, which support extracting the ingredients for two corresponding kinds of semantic-preserving code transformations (i.e., equivalent structure transformations and identifier renaming transformations).
With the guidance of reducing their code differences, the target input could be effectively transformed to a fault-revealing example based on the two kinds of transformations.
% As equivalent structure transformations do not affect the naturalness of generated examples, \tech{} first applies this kind of transformations to reduce code differences for improving the attack effectiveness (Section~\ref{sec:Structure}).
As equivalent structure transformations do not affect the naturalness of generated examples, \tech{} first applies this kind of transformations to reduce code differences for improving the generation of fault-revealing examples (Section~\ref{sec:Structure}).
% We introduce equivalent structure transformations and how to apply them to the target input for reducing the code differences in Section~\ref{sec:Structure}.
Then, we apply identifier renaming transformations to further reduce the code differences for improving the test effectiveness (Section~\ref{sec:remaning}).
% Here, \tech{} measures the semantic similarity between identifiers
% and designs an iterative process for applying this kind of transformations, so as 
% to guarantee the naturalness of generated examples.

Figure~\ref{fig:overview} shows the overview of \tech{}.
% In a nutshell, by successively applying equivalent structure transformations and identifier renaming transformations to the target input with the ingredient space defined by code differences between the selected reference inputs and the target one, 
% adversarial examples can be generated towards the direction of reducing code differences without altering the original semantics.
% the target input can gradually have smaller code differences with the reference inputs without altering the original semantics.
% In this way, the prediction result of the target input is more likely to be changed, leading to a successfully-attacking adversarial example.
% Due to the smaller ingredient search space (but including effective ingredients) and clearer attack direction (towards the direction of reducing code differences without altering the original semantics), the attack effectiveness could be improved by \tech{}. 
Due to the smaller ingredient search space (but including effective ingredients) and clearer generation direction (towards the direction of reducing code differences without altering the original semantics), the test effectiveness could be improved by \tech{}.

\begin{table}[]
\caption{Descriptions of equivalent structure transformations}
% \vspace{-2mm}
\label{tab:code_style}
\centering
\tabcolsep=2.1mm
\begin{adjustbox}{max width=0.5\textwidth,center}
    \begin{tabular}{ ll }
    \toprule
    \textbf{Transformation} & \textbf{Description} \\ \midrule
    \multirow{2}{*}{$R_1$-$loop$} & equivalent transformation among \texttt{for} structure \\
    & and \texttt{while} structure \\
    \midrule
    \multirow{2}{*}{$R_2$-$branch$} & equivalent transformation between \texttt{if-else(-if)} \\
    & structure and \texttt{if-if} structure \\
    \midrule
    \multirow{2}{*}{$R_3$-$calculation$} & 
    equivalent numerical calculation transformation, e.g., \\
    & \texttt{++}, \texttt{--}, \texttt{+=}, \texttt{-=}, \texttt{*=}, \texttt{/=}, \texttt{\%=}, \texttt{<<=}, \texttt{>>=}, \texttt{\&=}, \texttt{|=} , \texttt{ $\hat{}$ = } \\
    \midrule
    \multirow{2}{*}{$R_4$-$constant$} & 
    equivalent transformation between a constant and \\
    & a variable assigned by the same constant \\
    \bottomrule
    \end{tabular}
\end{adjustbox}
% \vspace{-5mm}
\end{table}

% \vspace{-1mm}
\subsection{Reference Inputs Selection}
\label{sec:ReferenceInputsSelection}
The goal of reference inputs is to reduce the ingredient space.
The reduced space should include the ingredients that are effective to transform the target input to a fault-revealing example.
% In this way, the adversarial attack process can be largely improved by searching for effective ingredients more efficiently.
In this way, the adversarial example generation process can be largely improved by searching for effective ingredients more efficiently.
% Although all the inputs that have different prediction results with the target one can provide ingredients for altering the prediction result of the target one after transformations, their capabilities for successful attacks can be different.
Although all the inputs that have different prediction results with the target one can provide ingredients for altering the prediction result of the target one after transformations, their capabilities for adversarial example generation can be different.
To transform the target input to a fault-revealing example with fewer perturbations,
% \tech{} should select the reference inputs, which provide the ingredients that are more likely to conduct successful attacks on the target input.
\tech{} should select the reference inputs, which provide the ingredients that are more likely to generate a fault-revealing example on the target input.
Similar to the existing work~\cite{sharif2016accessorize,grosse2017statistical,prakash2018deflecting,shen2020multiple}, we assume the prediction result of the target input is more likely to be changed from its original class denoted as $c_i$ (with the largest probability predicted by the target model) to the class with the second largest probability (denoted as $c_j$).
Hence, the ingredients in the inputs belonging to $c_j$ are more likely to generate a fault-revealing example on the target input, and thus \tech{} selects the inputs belonging to $c_j$ as the initial set of reference inputs.
Note that all the reference inputs are selected from the training set to (1) avoid introducing the contents beyond the cognitive scope of the model and (2) ensure the sufficiency of the inputs belonging to $c_j$.
% since the ground-truth label of each training input is known, 
Moreover, we only consider the training inputs whose prediction results are consistent with their ground-truth labels so as to avoid introducing noise. 
% \jj{noise is unclear.}. \tz{Moreover, analyzing perturbations and robustness upon those incorrectly handled examples is not quite meaningful~\cite{zhang2022towards}.
% Therefore, we only consider the training inputs whose prediction results are consistent with their ground-truth labels.}
% We denote the target input as $x$, and can obtain its confidence vector (denoted as $P=\{p_1,\ldots,p_n\}$) predicted by the target deep code mode (denoted as $M$ with $n$ classes $C=\{c_1,\ldots,c_n\}$), where $p_i$ ($1\leq i\leq n$) is the probability that $x$ is predicted as $c_i$.

However, the number of inputs belonging to the same class (i.e., $c_j$ as above) could be large, and thus the ingredient space constituted by code differences between them and the target input could be also large.
Hence, to further reduce the ingredient space for more effective adversarial example generation, \tech{} selects a subset of inputs with high similarity to the target input from the initial set of reference inputs, as the final set of reference inputs used by \tech{}.
This is because smaller code differences can effectively limit the number of ingredients, leading to smaller ingredient space.
% \tech{} does not select only one reference input, as too small ingredient space could incur a high risk of missing too many ingredients contributing to successful attacks.
\tech{} does not select only one reference input, as too small ingredient space could incur a high risk of missing too many ingredients contributing to fault-revealing example generation.
% That is, \tech{} selects a small set of reference inputs following the above two steps of selection to balance the ingredient space size and the amount of ingredients contributing to successful attacks in the space.

We further introduce how to measure the similarity between the target input (denoted as $t$) and a reference input (denoted as $r$) for the above selection.
In general, we can adopt some pre-trained models to represent the code as a vector and then measure code similarity by calculating the vector distance, like many existing studies~\cite{zhao2018deepsim,alon2019code2vec,feng2020codebert}.
% However, as presented in Section~\ref{sec:overview}, \tech{} first applies equivalent structure transformations (rather than identifier renaming transformations) to reduce code differences for adversarial attacks.
However, as presented in Section~\ref{sec:overview}, \tech{} first applies equivalent structure transformations (rather than identifier renaming transformations) to reduce code differences for adversarial example generation.
% , as this kind of transformations does not affect the naturalness of generated examples.
Moreover, the identifiers used in different code snippets are usually different due to the enormous identifier space, which may lead to the low similarity between various code snippets.
Hence, when measuring code similarity, \tech{} eliminates the influence of identifiers by replacing them with the placeholder {\tt <unk>}.
Specifically, \tech{} first represents $t$ and $r$ after placeholder replacement as vectors respectively based on CodeBERT~\cite{feng2020codebert} (one of the most widely-used pre-trained models~\cite{zhou2021assessing,lu2021codexglue,pan2021empirical}), and then calculates the cosine similarity between vectors.
% As the descending order of the calculated similarity, \tech{} selects Top-$N$ reference inputs for the follow-up attack process.
As the descending order of the calculated similarity, \tech{} selects Top-$N$ reference inputs for the follow-up generation process.
Note that to make the selection process efficient, we randomly sampled $U$ inputs from the initial set for this step of selection.
We will study the influence of both $U$ and $N$ on \tech{} in Section~\ref{sec:dis}.
\subsection{Equivalent Structure Transformation}
\label{sec:Structure}
Based on the small set of selected reference inputs and the guidance of code differences, 
% \tech{} then extracts ingredients from the space defined by their brought code differences over the target input. 
% \tz{not so clear to read}.\yl{Is "space" really necessary?}
% That is, 
% \tech{} transforms the target input to an adversarial example
% \yl{Here, CODA does not know if the transformed inputs will lead to a success attack, therefore, successfully-attacking example is inappropriate. Perhaps "candidates" is better.} 
% towards the direction of reducing code differences between them and the target input.
\tech{} first reduces structure differences by applying equivalent structure transformations to the target input. 
% as they do not affect the naturalness of generated examples.

To preserve the semantics of the target input, we design four categories of equivalent structure transformations (without affecting code naturalness) in \tech{} following the existing work in metamorphic testing and code refactoring~\cite{nakamura2016random,cheers2019spplagiarise,henke2022semantic}.
In particular, we systematically consider all common kinds of code structures, i.e., \textit{loop} structures, \textit{branch} structures, and \textit{sequential} structures (including numerical calculation and constant usage).
We explain the four categories in detail in Table~\ref{tab:code_style}.
% , each of which is also illustrated with an example.\tz{We explain the four categories in detail in Table~\ref{tab:code_style}.}
For each category of transformations, it may include several specific rules. 
% \tz{For each category of transformations in Table~\ref{tab:code_style}, it may include several specific rules.}\jj{why revise it?}
For example, the rules of transformation on {\tt +=} and transformation on {\tt -=} belong to the category of \textit{$R_3$-calculation}, and the rules of transforming {\tt for} loop to {\tt while} loop and transforming {\tt while} loop to {\tt for} loop belong to the category of \textit{$R_1$-loop}.
In total, \tech{} has 20 specific rules for the four categories of transformations.
Due to the space limit, we list all these specific rules at our project homepage~\cite{coda2022}.
% \jj{what's the total number of specific rules.} \tz{20}
These rules are general to the vast majority of popular programming languages (e.g., C, C++, and Java), which ensures the generality of \tech{} to a large extent.
% , but there are some rules that may be inapplicable to some specific programming languages.
% Please note that not all the rules are applicable to the code programmed by any programming language.
% For example, {\tt ++} and {\tt --} in \textit{$R_3$-calculation} are not supported by Python.
Note that in \textit{$R_4$-constant}, the newly-defined variable cannot be the same as the existing variables in the code; otherwise, it may incur grammar errors and alter the original semantics.

Then, we illustrate how to apply these rules 
% guided by code differences.
for reducing code differences.
Each rule involves two structures, i.e., the one before transformation ($s_b$) and the one after transformation ($s_a$).
\tech{} first counts the occurring times of $s_b$ and $s_a$ in the set of selected reference inputs (denoted as $n_b$ and $n_a$), and then calculates their occurring distribution, i.e., $\frac{n_b}{n_b+n_a}$ and $\frac{n_a}{n_b+n_a}$.
Further, \tech{} applies each rule in a probabilistic way to reduce the occurring distribution differences in terms of $s_b$ and $s_a$ between reference inputs and the target input since probabilistic methods tend to improve effectiveness compared to deterministic methods in general~\cite{yang2021semi}.
In this way, the structure differences in terms of $s_b$ and $s_a$ can be reduced.
Specifically, for each occurrence of $s_b$ in the target input, \tech{} applies this rule with the probability of $\frac{n_a}{n_b+n_a}$, also indicating it can be retained with the probability of $\frac{n_b}{n_b+n_a}$.

In this step, \tech{} obtains $M$ inputs from the target input, each of which is generated by applying all the applicable rules together on the target input in the above probabilistic way, and then selects the input with the highest average similarity (also measured by the method described in Section~\ref{sec:ReferenceInputsSelection}) to the selected reference inputs as the one for the follow-up generation process.

\subsection{Identifier Renaming Transformation}
\label{sec:remaning}
To facilitate the generation of fault-revealing examples, 
\tech{} then applies identifier renaming transformations to further reduce code differences.
Inspired by the existing work~\cite{zhang2020generating,zhang2022towards,yang2022natural}, 
% With the guidance of code differences,
identifier renaming transformation in \tech{} refers to replacing the name of an identifier in the target input with the name of an identifier in the selected reference inputs.
For ease of presentation, we denote the set of identifiers in the target input as $V_t$ and the set of identifiers in the selected reference inputs as $V_r$.
To preserve the semantics of the target input and ensure the grammatical correctness of the generated example, \tech{} ensures that the identifiers used for replacement do not exist in the target input.

Then, we illustrate how to apply this kind of transformations to the input obtained from the last step.
As demonstrated by the existing work~\cite{zhang2020generating,zhang2022towards,yang2022natural}, renaming identifiers is effective to generate fault-revealing examples, but can negatively affect naturalness of generated examples.
To ensure the naturalness of generated examples, \tech{} considers the semantic similarity between identifiers and designs an iterative transformation process like ALERT~\cite{yang2022natural}.
Specifically, \tech{} measures the semantic similarity between each identifier in $V_t$ and each identifier in $V_r$ by representing each identifier as a vector via word embedding.
Here, \tech{} builds the pre-trained language model with FastText~\cite{bojanowski2017enriching} and calculates the cosine similarity between vectors to measure their semantic similarity.
% \yl{Report the similarity of identifier pairs in motivating example.}
Then, \tech{} prioritizes each pair of identifiers in the descending order of their semantic similarity, and iteratively applies this transformation based on each pair of identifiers in the ranking list, which ensures more natural transformations can be first performed.
\tech{} ensures the pair of identifiers will not introduce repetitive identifiers in the generated example in each iteration; otherwise, this pair will be discarded.
After each iteration, \tech{} invokes the target model to check whether a fault is detected by the generated example.
% a fault-revealing example is generated.
% \textit{The iterative attack process terminates until a successfully-attacking adversarial example is generated or all the pairs are used by this transformation.}

Following the existing work~\cite{yang2022natural,zhang2022towards},
the iterative generation process terminates until a fault-revealing example from the target input is generated or all the pairs are used by this transformation.
This is because more fault-revealing examples generated from the same target input tend to detect duplicate faults in the model.
The setting can be adjusted by users according to the demand in practice.
In this work, we directly adopt the setting from the existing work~\cite{yang2022natural,zhang2022towards} in \tech{}.
%yang2022natural,zhang2022towards
% \jj{That is, for each target input, \tech{} can generate \textbf{at most one} successfully-attacking adversarial example, which shares the same setting as the state-of-the-art techniques (i.e., ALERT and CARROT).}\jj{to move}

Overall, \tech{} only invokes the target model when checking if a fault is detected by the generated example, which is necessary for testing the model.
Hence, \tech{} can largely reduce the number of model invocations compared with the existing techniques, confirmed by our study
(Section~\ref{sec:rq1}).

\section{Evaluation Design}
\label{sec:design}
In the study, we address four research questions (RQs):

\begin{itemize}
    \item \textbf{RQ1:} How does \tech{} perform in terms of effectiveness and efficiency compared with state-of-the-art techniques?
    
    \item \textbf{RQ2:} Are the adversarial examples generated by \tech{} useful to enhance the robustness of deep code models?
    
    \item \textbf{RQ3:} Does each main component contribute to the overall effectiveness of \tech{}?

    \item \textbf{RQ4:} Are the adversarial examples generated by \tech{} natural for humans?
\end{itemize}

% In this section, we conduct extensive research to sufficiently evaluate the performance of our proposed adversarial attack technique, i.e., \tech{}.
% Specifically, we evaluate \tech{} from four aspects, including effectiveness and efficiency in neural attack -- \textbf{RQ1}, naturalness of adversarial examples -- \textbf{RQ2}, enhancements to deep code models -- \textbf{RQ3}, and contribution of each component -- \textbf{RQ4}.

\begin{table}[t]
\caption{Statistics of our used subjects}
% \vspace{-2mm}
\label{tab:tasks_and_models}
\centering
\tabcolsep=1.2mm
\begin{adjustbox}{max width=.49\textwidth,center}
    \begin{threeparttable}
        \begin{tabular}{ llclll }
        \toprule
        \textbf{Task} & \textbf{Train/Val/Test} & \textbf{Class} & \textbf{Language} & \textbf{Model} & \textbf{Acc.} \\ \midrule
        \multirow{3}{*}{\makecell[l]{Vulnerability \\ Prediction}} & \multirow{3}{*}{ 21,854/2,732/2,732 } & \multirow{3}{*}{2} & \multirow{3}{*}{C} & CodeBERT & 63.76\% \\
        &&&& GCBERT & 63.65\% \\
        &&&& CodeT5 & 63.83\% \\
        \hline
        \multirow{3}{*}{\makecell[l]{Clone \\ Detection}} & \multirow{3}{*}{ 90,102/4,000/4,000 } & \multirow{3}{*}{2} & \multirow{3}{*}{Java} & CodeBERT & 96.97\% \\
        &&&& GCBERT & 97.36\% \\
        &&&& CodeT5 & 98.08\% \\
        \hline
        \multirow{3}{*}{\makecell[l]{Authorship \\ Attribution}} & \multirow{3}{*}{ 528/–/132 } & \multirow{3}{*}{66} & \multirow{3}{*}{Python} & CodeBERT & 90.35\% \\
        &&&& GCBERT & 89.48\% \\
        &&&& CodeT5 & 92.30\% \\
        \hline
        \multirow{3}{*}{\makecell[l]{Functionality \\ Classification}} & \multirow{3}{*}{ 41,581/–/10,395 } & \multirow{3}{*}{104} & \multirow{3}{*}{C} & CodeBERT & 98.18\% \\
        &&&& GCBERT & 98.66\% \\
        &&&& CodeT5 & 98.79\% \\
        \hline
        \multirow{3}{*}{\makecell[l]{Defect \\ Prediction}} & \multirow{3}{*}{ 27,058/–/6,764 } & \multirow{3}{*}{4} & \multirow{3}{*}{C/C++} & CodeBERT & 84.37\% \\
        &&&& GCBERT & 83.98\% \\
        &&&& CodeT5 & 81.54\% \\
        \bottomrule
        \end{tabular}
        \begin{tablenotes}
            \footnotesize
            \item[]* GCBERT is short for GraphCodeBERT.
        \end{tablenotes}
    \end{threeparttable}
\end{adjustbox}
% \vspace{-5mm}
\end{table}

% \vspace{-2mm}
\subsection{Subjects}
\label{sec:sub}

\subsubsection{Datasets and Tasks}
\label{sec:task}
% \jj{we may add the programming language information in the dataset, which can show our approach (especially our structure transformations) can be generalized to any programming languages.} \tz{done}
To sufficiently evaluate \tech{}, we consider all the five code-based tasks in the studies of evaluating state-of-the-art techniques (i.e., CARROT~\cite{zhang2022towards} and ALERT~\cite{yang2022natural}).
The statistics of datasets are shown at the first four columns in Table~\ref{tab:tasks_and_models}, each of which represents the task, the number of inputs in the training/validation/test set, the number of classes for the classification task, and the programming language for the inputs.
% , the used pre-trained DL model for solving the task (to be explained in Section~\ref{sec:model}), the accuracy of the model fine-tuned on the dataset, respectively.
% We briefly introduce the five tasks and the corresponding dataset as follows.
% They are vulnerability prediction, clone detection, authorship attribution, functionality classification, and defect prediction.

% \tz{maybe we can delete some description?}
The task of \textit{vulnerability prediction} aims to predict whether a given code snippet has vulnerabilities.
Its used dataset is extracted from two C projects~\cite{zhou2019devign}. 
% (i.e., FFmpeg~\cite{ffmpeg2022} and Qemu~\cite{qemu2022}) by Zhou et al. 
% and has been integrated as part of the CodeXGLUE benchmark~\cite{lu2021codexglue}.
% Each input is a function and labeled as either containing vulnerabilities or not.
% The goal of vulnerability prediction is to predict whether code snippets contain vulnerabilities. 
% The dataset for this task is part of the CodeXGLUE benchmark~\cite{lu2021codexglue}, and a total of 27318 code snippets are labeled as either containing vulnerabilities or clean. 
The task of \textit{clone detection} aims to detect whether two given code snippets are equivalent in semantics.
Its used dataset is from BigCloneBench~\cite{svajlenko2014towards}, the most widely-used dataset for clone detection.
% The existing work~\cite{yang2022natural} 
% % filtered the data without labels and balanced the numbers of positive and negative instances, and then 
% randomly sampled 90,102/4,000/4,000 inputs from the benchmark for training/validation/testing, to make the experiment at a computationally friendly scale.
% In our study, we used the same dataset.
% Following the settings of prior researches~\cite{wang2020detecting, wei2017supervised}, we randomly selected 98,102 datasets with labels from BigCloneBench~\cite{svajlenko2014towards}, a widely used clone detection benchmark. 
The task of \textit{authorship attribution} aims to identify the author of a given code snippet.
Its used dataset is the Google Code Jam (GCJ) dataset~\cite{alsulami2017source}, which contains 660 Python code files and 66 author information after removing 40 code files in other programming languages~\cite{yang2022natural}.
% The experimental dataset comes from the Google Code Jam (GCJ) dataset~\cite{alsulami2017source}, which contains 660 code files and 66 author information after filtering.
The task of \textit{functionality classification} aims to classify the functionality of a given code snippet.
% If code snippets solve the same problem, they are regarded to have the same functionality~\cite{zhang2019novel}. 
Its used dataset is the Open Judge (OJ) benchmark~\cite{mou2016convolutional}.
% , which has been also integrated as part of the CodeXGLUE benchmark~\cite{lu2021codexglue}.
The task of \textit{defect prediction} aims to predict whether a given code snippet is defective and its defect type.
Its used dataset is the CodeChef dataset~\cite{codechef2022}.
% , which is labeled by the execution results on the CodeChef platform. 
% (i.e., four defect types: no defect, wrong answer, timeout, runtime error).

\subsubsection{Models}
\label{sec:model}
Same as the existing work~\cite{yang2022natural}, we used the state-of-the-art pre-trained models, i.e., CodeBERT~\cite{feng2020codebert} and GraphCodeBERT~\cite{guo2021graphcodebert}, in our study, and also used the more recent CodeT5~\cite{wang2021codet5}.
% There are many other pre-trained models in the experiment of CARROT (such as, LSTM and TBCNN), but they do not show
% superior performance than CodeBERT, GraphCodeBERT, and CodeT5, so we
% investigate the latter three in this work.
Several alternative pre-trained models have been studied in the experiments of CARROT (such as LSTM and TBCNN). 
However, these models did not demonstrate superior performance compared to CodeBERT, GraphCodeBERT, and CodeT5. 
As a result, our investigation in this study focused on the latter three models.

% Following the existing work~\cite{yang2022natural}, we adopted three state-of-the-art pre-trained models for code-based tasks (i.e., CodeBERT~\cite{feng2020codebert}, GraphCodeBERT~\cite{guo2021graphcodebert}, and CodeT5~\cite{wang2021codet5}), and then 
We fine-tuned them on the five tasks based on the corresponding datasets, respectively.
When fine-tuning CodeBERT and GraphCodeBERT on these tasks (except GraphCodeBERT on functionality classification and defect prediction), we used the same hyper-parameter settings provided by the existing work~\cite{yang2022natural,zhang2022towards}.
As there is no instruction on the hyper-parameter settings for fine-tuning GraphCodeBERT on functionality classification and defect prediction, we used the same settings as the one used by authorship attribution (they are all multi-class classification tasks).
Indeed, the achieved model performance outperforms that achieved by the models (e.g., TBCNN~\cite{mou2016convolutional} and CodeBERT~\cite{feng2020codebert}) used in the existing work~\cite{zhang2022towards} on the same datasets~\cite{zhang2022towards}, indicating that the transferred hyper-parameter settings are reasonable.
Similarly, we used the same settings for CodeT5 as CodeBERT and the achieved performance of CodeT5 is better than (or close to) those achieved by CodeBERT and GraphCodeBERT on all the tasks.
The detailed settings can be found at our project homepage~\cite{coda2022}.
In total, we obtained 15 deep code models as the subjects.
The last two columns in Table~\ref{tab:tasks_and_models} show the used pre-trained model and the accuracy of the deep code model after fine-tuning on the corresponding task, respectively.
Overall, the subjects used in our study are diverse, involving different tasks, different pre-trained models, different numbers of classes, different programming languages, etc.
It is very helpful to sufficiently evaluate the performance of \tech{}.

% In particular, since there is no specification on the hyper-parameter settings for fine-tuning GraphCodeBERT for functionality classification and defect prediction, 
% we use the same settings as other tasks and obtain slightly higher performance than all models in CARROT.
% This indicates that all the target models used in our experiments are sufficiently fine-tuned.

% \vspace{-1mm}
\subsection{Compared Techniques}
\label{sec:techniques}
In the study, we compared \tech{} with two state-of-the-art adversarial example generation techniques for deep code models, i.e., \textbf{CARROT}~\cite{zhang2022towards} and \textbf{ALERT}~\cite{yang2022natural}, which have been introduced in Section~\ref{sec:intro} (the third paragraph).
We adopted their implementations and the recommended parameter settings provided by the corresponding papers~\cite{zhang2022towards, yang2022natural}.
% As the original version of CARROT can only support to attack C/C++ code, we extended it to attack Python and Java code for sufficient comparison.
As the original version of CARROT can only support C/C++ code, we extended it to Python and Java code for sufficient comparison.

% In our study, We choose two state-of-the-art algorithms, \textbf{CARROT}~\cite{zhang2022towards} and \textbf{ALERT}~\cite{yang2022natural}, as our baseline neural attack algorithms for comparison. 
% CARROT is an effective adversarial attack technique against DL models with gradient-guided source code processing. 
% Besides, CARROT designed two levels of attacks, namely token-level and statement-level. 
% The original CARROT can only attack C/C++ programs, so we extend it to attack Python and Java code snippets.
% ALERT is able to generate natural substitutes, which are natural adversarial codes for humans. 
% Besides, ALERT utilizes greedy and genetic algorithms to efficiently discover adversarial examples. 
% Regarding them, we used the hyper-parameters recommended by their original papers~\cite{zhang2022towards, yang2022natural}.

% \begin{table}[]
% \caption{Hyper-parameters of the target models}
% \label{tab:hyper-parameters}
% \centering
% \tabcolsep=4.0mm
% \begin{adjustbox}{max width=.50\textwidth,center}
%     \begin{tabular}{ lcc }
%     \toprule
%     \textbf{Hyper-parameters} & \textbf{CodeBERT} & \textbf{GraphCodeBERT} \\ \midrule
%     Input length & 512 & 512 \\
%     Batch size & 16 & 64 \\
%     Max epoch & 30 & 30 \\
%     Optimizer & Adam & Adam \\
%     Max norm & 1.0 & 1.0 \\
%     Learning rate & 5e-5 & 2e-5 \\
%     Early stopping & $\checkmark$ & $\checkmark$ \\
%     $N_1$ & 256 & 256 \\
%     $N_2$ & 64 & 64 \\
%     \bottomrule
%     \end{tabular}
% \end{adjustbox}
% \vspace{-4mm}
% \end{table}

% \vspace{-1mm}
\subsection{Implementations}
\label{sec:implementations}

We implemented \tech{} in Python and adopted tree-sitter~\cite{treesitter2022} to extract identifiers from code following the existing work~\cite{yang2022natural}.
We set the parameters in \tech{} by conducting a preliminary experiment, i.e., $U=256$ (the number of sampled inputs from the initial set for similarity calculation), $N=64$ (the number of reference inputs selected after similarity calculation), and $M=64$ (the number of examples generated via structure transformations).
We will discuss the influence of the settings of main parameters in Section~\ref{sec:dis}.
% We determined the configuration of hyper-parameters in \tech{} according to the performance of the validation set and previous work, and the specific hyper-parameters settings are shown in Table~\ref{tab:hyper-parameters}. 
% And we implemented a identifier extractor based tree-sitter~\cite{treesitter2022}, which can extract all identifier names from code snippets.
% The attack process terminates until a successfully-attacking adversarial example is generated or all the pairs are used by this transformation.
% With our implementation, researchers and practitioners can directly use it to build adversarial examples for any given code snippet. 
All the experiments were conducted on a server with an Ubuntu 20.04 system with Intel(R) Xeon(R) Silver 4214 @ 2.20GHz CPU, and NVIDIA GeForce RTX 2080 Ti GPU.

% \subsection{Process}
% \label{sec:process}
% \jj{describe the experimental process.}
% \vspace{-1mm}
% \vspace{-1.5mm}
\section{Results and Analysis}
\label{sec:results}
% \vspace{-1.0mm}
 \subsection{RQ1: Effectiveness and Efficiency }
\label{sec:rq1}

\begin{table*}[t]
    \caption{Effectiveness comparison in terms of RFR}
    % \vspace{-2mm}
    \label{tab:asr}
    \centering
    \tabcolsep=4.0mm
    \begin{adjustbox}{max width=1.0 \textwidth,center}
        \begin{tabular}{ cccccccccc }
            \toprule
        	\multirow{2}{*}{\textbf{Task}} & \multicolumn{3}{c}{\textbf{CodeBERT}} & \multicolumn{3}{c}{\textbf{GraphCodeBERT}} & \multicolumn{3}{c}{\textbf{CodeT5}} \\ \cmidrule(lr){2-4} \cmidrule(lr){5-7} \cmidrule(lr){8-10}
        	& \textbf{CARROT} & \textbf{ALERT} & \textbf{\tech{}} & \textbf{CARROT} & \textbf{ALERT} & \textbf{\tech{}} & \textbf{CARROT} & \textbf{ALERT} & \textbf{\tech{}} \\
        	\midrule
        	Vulnerability Prediction & 33.72\% & 53.62\% & \textbf{89.58\%} & 37.40\% & 76.95\% & \textbf{94.72\%} & 84.32\% & 82.69\% & \textbf{98.87\%} \\
        	Clone Detection & 20.78\% & 27.79\% & \textbf{44.65\%} & 3.50\% & 7.96\% & \textbf{27.37\%} & 12.89\% & 14.29\% & \textbf{42.07\%} \\
        	Authorship Attribution & 44.44\% & 35.78\% & \textbf{79.05\%} & 31.68\% & 61.47\% & \textbf{92.00\%} & 20.56\% & 66.41\% & \textbf{97.17\%} \\
        	Functionality Classification & 44.15\% & 10.04\% & \textbf{56.74\%} & 42.76\% & 11.22\% & \textbf{57.44\%} & 38.26\% & 35.37\% & \textbf{78.07\%} \\
        	Defect Prediction & 71.59\% & 65.15\% & \textbf{95.18\%} & 79.08\% & 75.87\% & \textbf{96.58\%} & 38.26\% & 35.37\% & \textbf{78.07\%} \\
        	\midrule
        	Average & 42.94\% & 38.48\% & \textbf{73.04\%} & 38.88\% & 46.69\% & \textbf{73.62\%} & 33.91\% & 40.99\% & \textbf{70.96\%} \\
            \bottomrule
        \end{tabular}
    \end{adjustbox}
    % \vspace{-4mm}
\end{table*}

% \begin{figure}[t]
%     \centering
%     \includegraphics[width=0.60\linewidth]{figs/unique_successfully_attacking.pdf}
%     % \vspace{-4mm}
%     \caption{ Comparison in terms of the number of unique successfully-attacking adversarial examples }
%     \label{fig:unique_successfully_attacking}
%     % \vspace{-6mm}
% \end{figure}

\begin{figure*}[t]
  % \vspace{-2mm}
  \centering
    \subfigure[PCD on testing CodeBERT]{\includegraphics[width=0.32\textwidth]{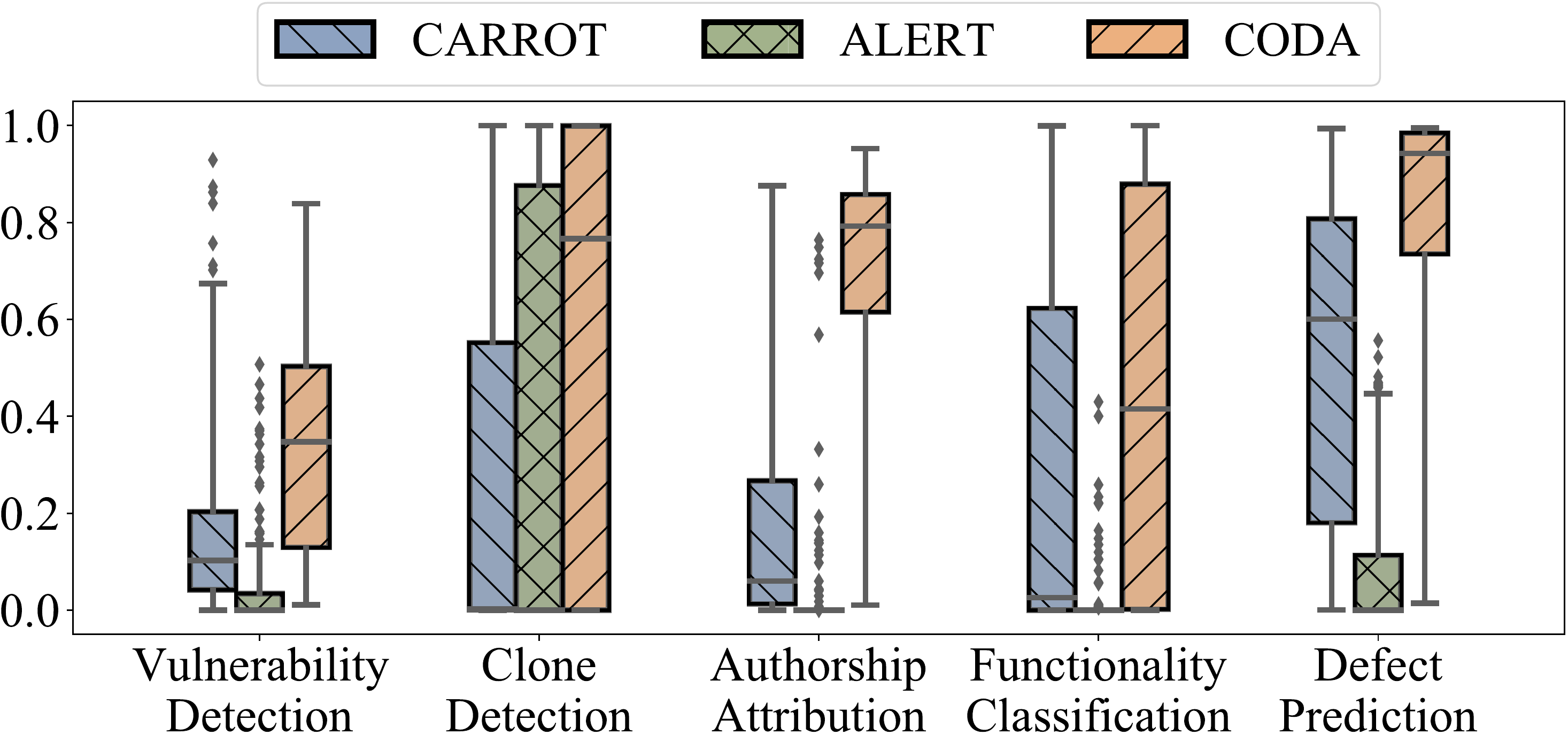}
    \label{fig:pcd_codebert}} 
    \subfigure[PCD on testing GraphCodeBERT]{\includegraphics[width=0.32\textwidth]{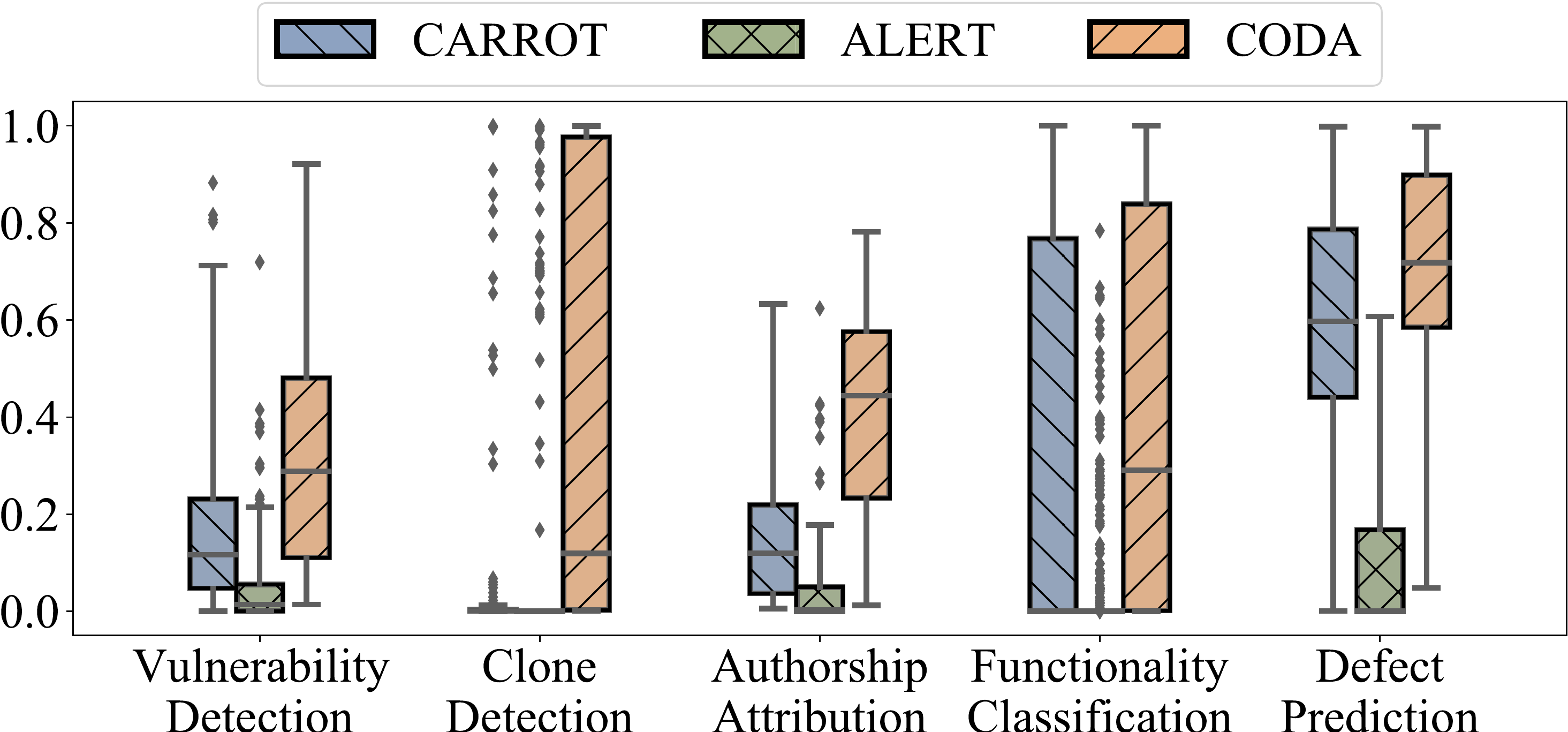}
    \label{fig:pcd_graphcodebert}}
    \subfigure[PCD on testing CodeT5]{\includegraphics[width=0.32\textwidth]{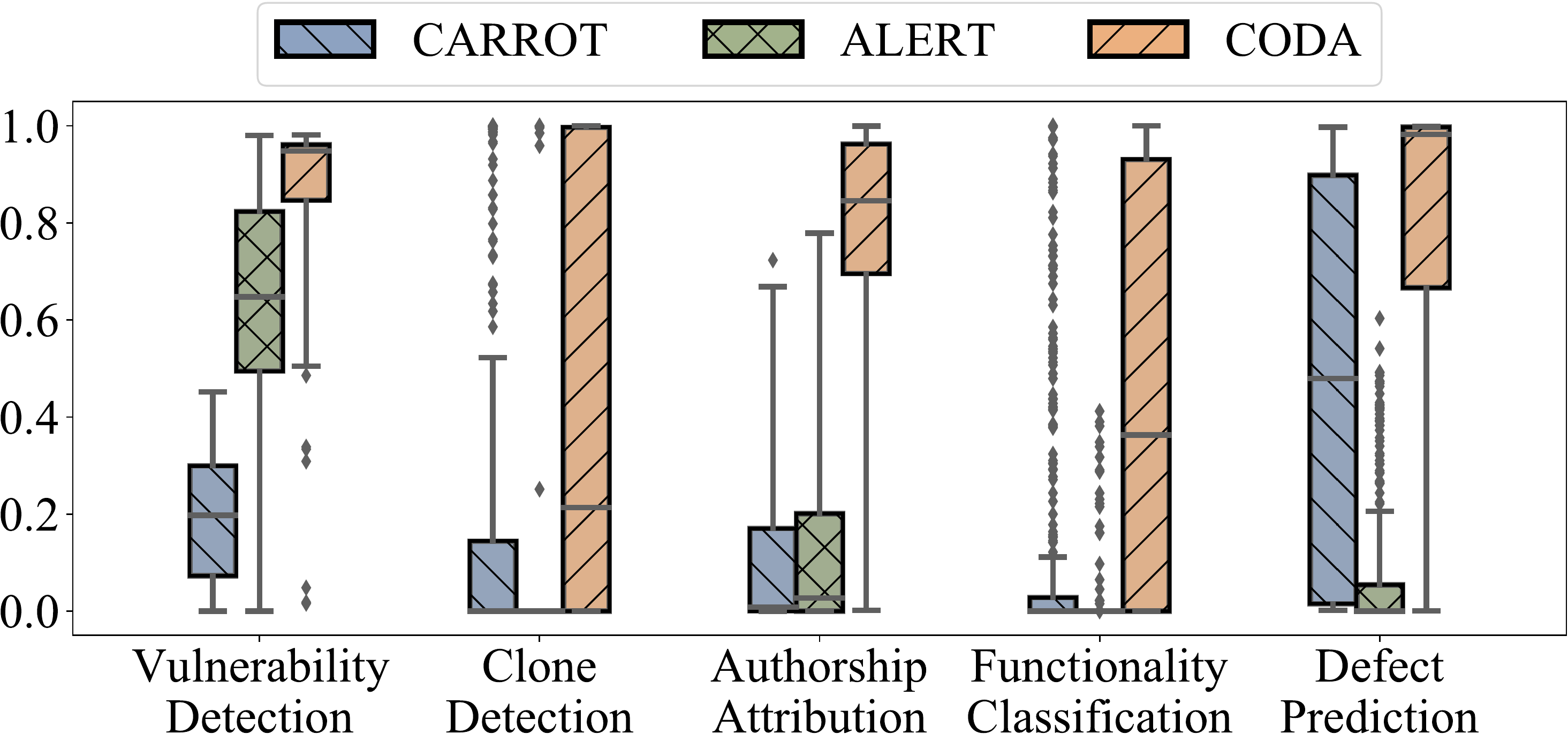}
    \label{fig:pcd_codet5}} \\
  % \vspace{-2mm}
  \caption{Comparison in terms of prediction confidence decrement}
  % \vspace{-2mm}
	\label{fig:rq1}
\end{figure*}

\begin{figure*}[!htb]
  % \vspace{-2mm}
  \centering
    \subfigure[Model invocations on testing CodeBERT]{\includegraphics[width=0.32\textwidth]{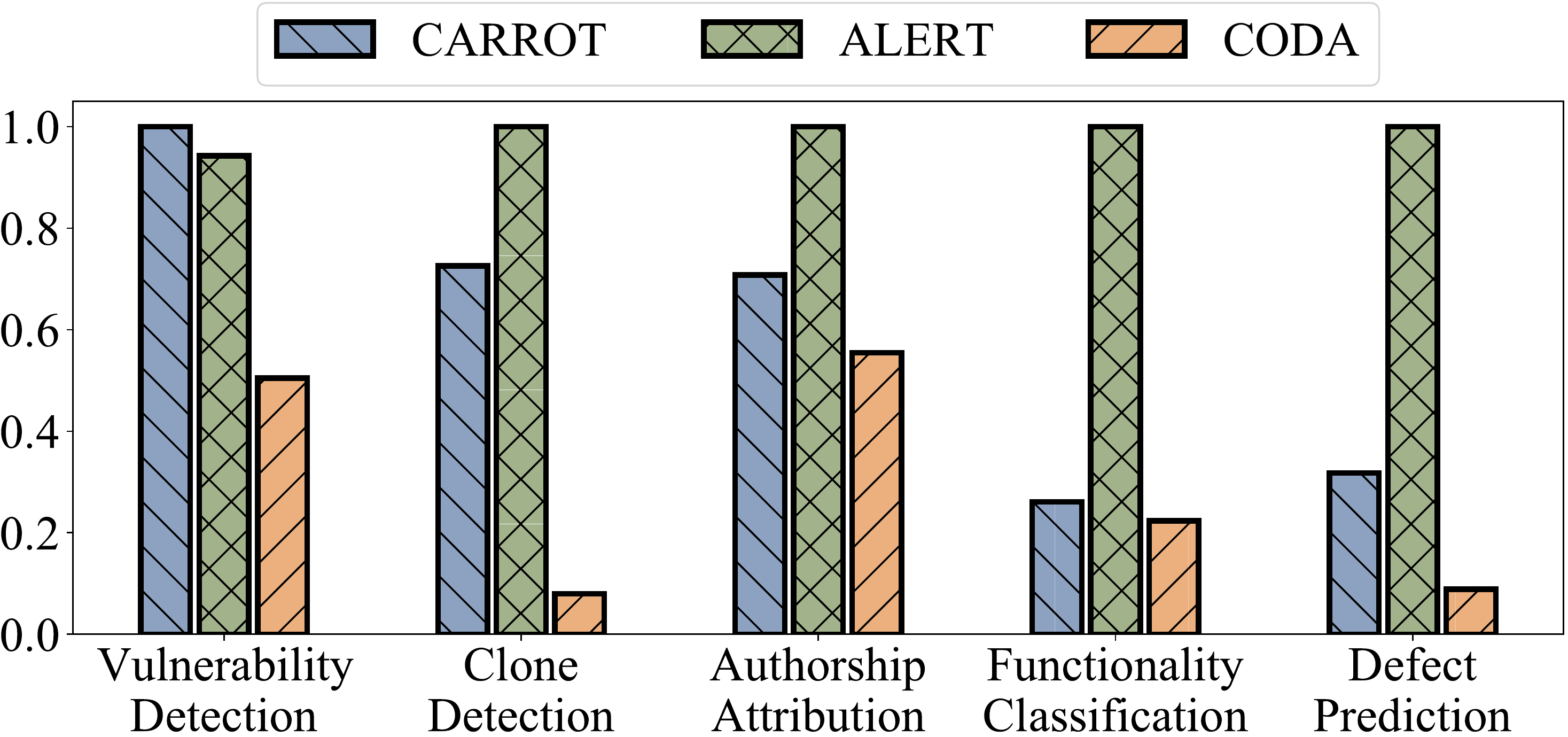}
    \label{fig:inv_codebert}} 
    \subfigure[Model invocations on testing GraphCodeBERT]{\includegraphics[width=0.32\textwidth]{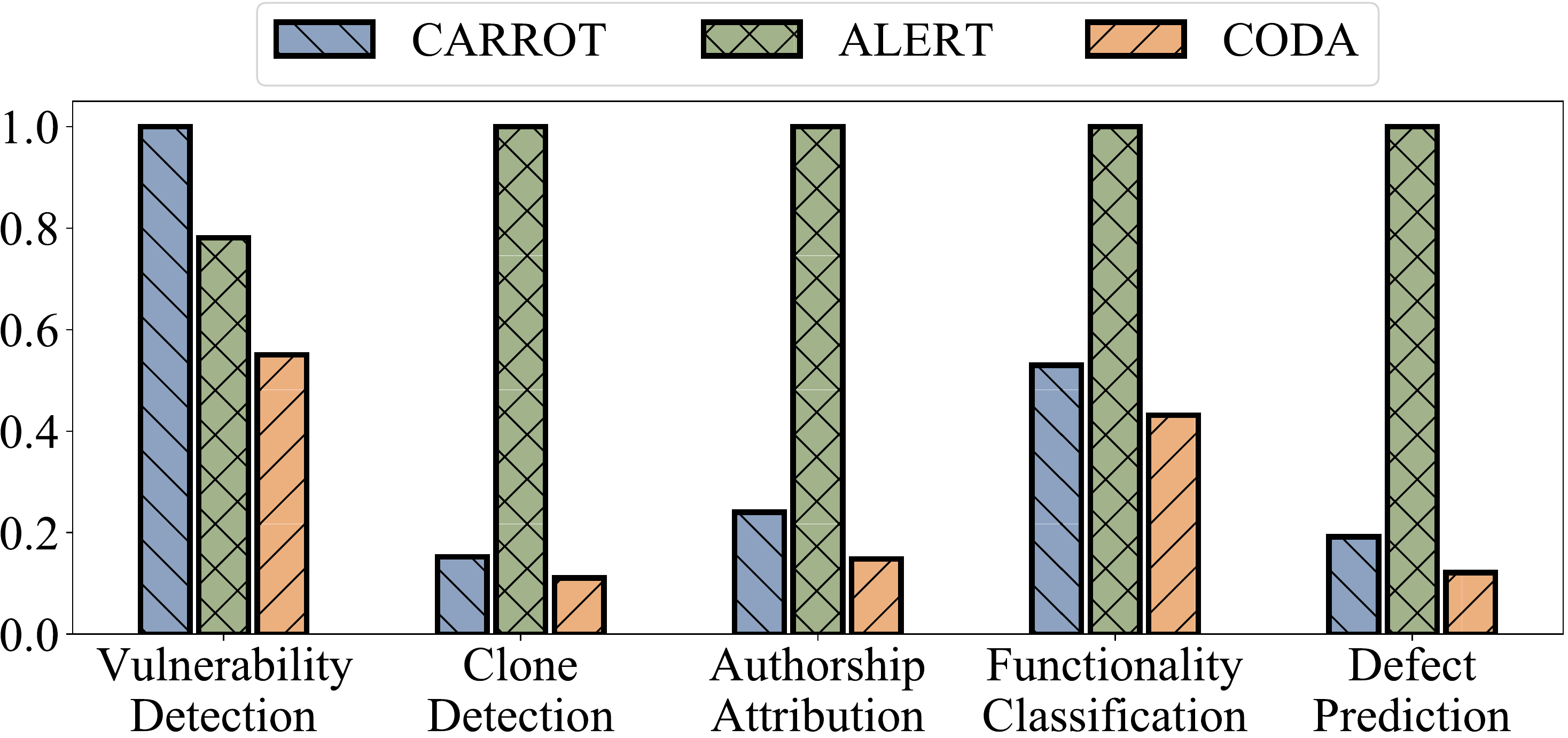}
    \label{fig:inv_graphcodebert}}
    \subfigure[Model invocations on testing CodeT5]{\includegraphics[width=0.32\textwidth]{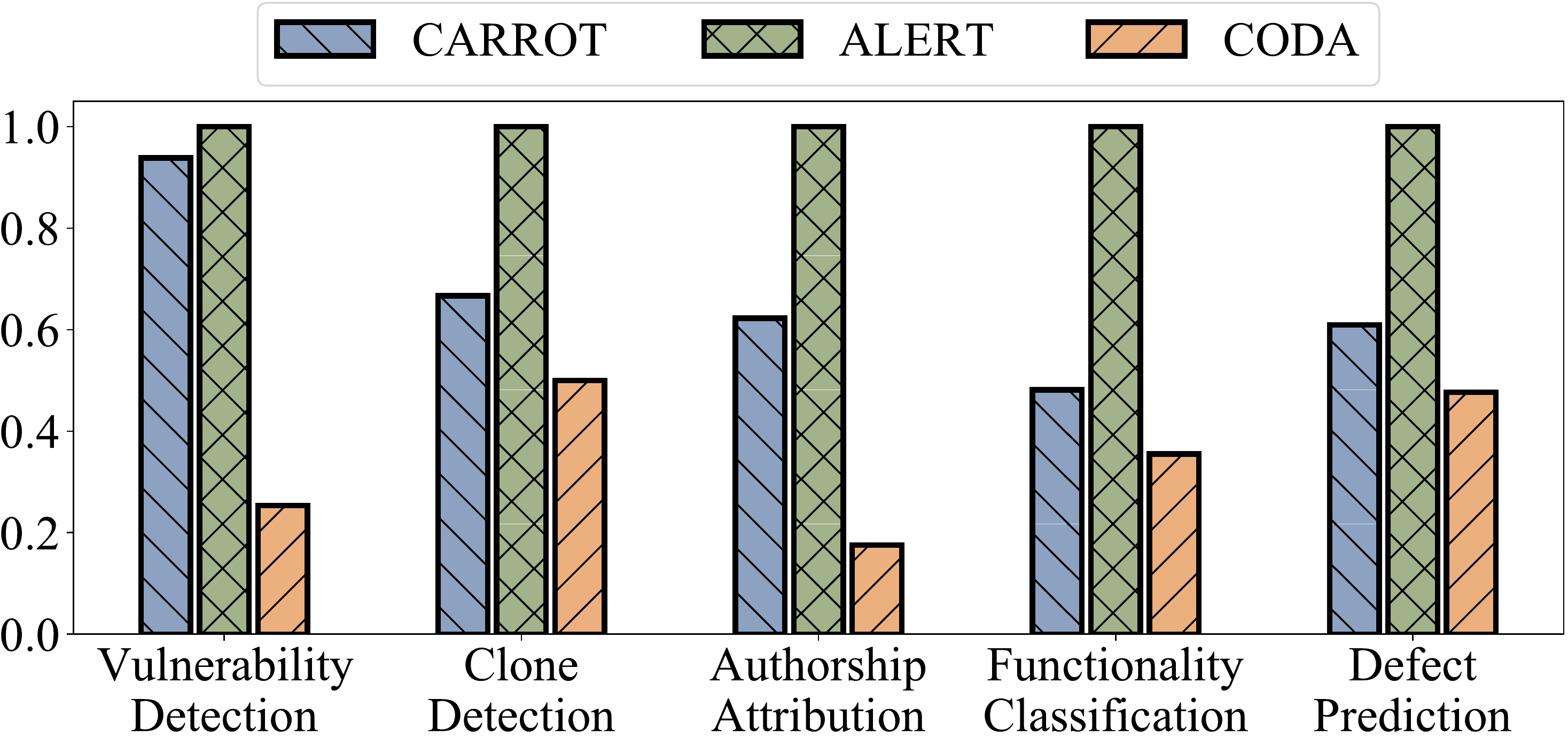}
    \label{fig:inv_codet5}} \\
	% \vspace{-0.1in}
  % \vspace{-2mm}
  \caption{Comparison in terms of model invocations (y-axis shows the normalized values following the existing work~\cite{yang2022natural})}
  % \vspace{-6mm}
	\label{fig:rq1}
\end{figure*}

\subsubsection{Setup}
For each deep code model, we applied \tech{}, CARROT, and ALERT to generate adversarial examples from each target input in the test set to test it, respectively.
We measured their effectiveness and efficiency based on the following metrics.
To reduce the influence of randomness, we repeated all the experiments (including those for other RQs) 10 times, and reported the average results.
% \tz{Moreover, we experimented ten times with \tech{} on all subjects and averaged the results to limit the effects of randomness.}

% Same as the attack process of ALERT and CARROT, \tech{} also produces \textit{at most one} successfully-attacking adversarial example for each target input.

We first measured the number of revealed faults by each technique.
As presented in Section~\ref{sec:remaning}, the faults revealed by several adversarial examples from the same target input tend to be duplicate as claimed in the existing work~\cite{gao2022adaptive,you2023regression}.
Hence, \tech{} produces \textit{at most one} fault-revealing example for each target input same as the generation process of ALERT and CARROT.
% Same as the generation process of ALERT and CARROT, \tech{} also produces \textit{at most one} inconsistency-revealing adversarial example for each target input.
That is, when a fault-revealing example is generated for a given target input, it will move to the next target input.
Therefore, the number of revealed faults is equal to the number of targets inputs from which a fault-revealing example is generated here.
% Therefore, following the existing work~\cite{yang2022natural}, we adopted the \textbf{attack success rate (A)SR (rather than the number of adversarial examples) to measure the effectiveness of each technique.
Since the sizes of different test sets are different, we reported \textbf{the rate of revealed faults (RFR)}, instead of the number of revealed faults, for better presentation, same as the existing work~\cite{gao2022adaptive,zhang2022towards}.
The rate of revealed faults for each subject refers to the ratio of the number of revealed faults to the total number of target inputs (that are correctly predicted as mentioned in Section~\ref{sec:problem}) in the test set of the subject.
% Hence, we measured the effectiveness of \tech{} in terms of \textbf{inconsistency-revealing rate (IRR)}. 
% IRR measures the ability of \tech{} to reveal inconsistencies, which is the ratio of the number of the target inputs from which an inconsistency-revealing adversarial example is generated to the total number of target inputs.
Larger RFR values mean better test effectiveness.

Also, it is important to measure whether the prediction confidence (i.e., the probability of being the ground-truth class of the target input) is decreased by the generated examples (although there is no fault-revealing example generated from a target input).
% \tz{
Reducing prediction confidence indicates that the generated examples make the model less robust.
% greater susceptibility of the examples to modifications in the predicted label, thereby facilitating the enhancement of the model~\cite{yang2022natural}.
% }
Hence, we calculated \textbf{prediction confidence decrement (PCD)} to measure the effectiveness of each technique.
PCD is calculated by the prediction confidence of the target input minus the minimum prediction confidence of the set of generated examples from the target input. 
If the former is smaller than the latter, we regard PCD to be 0, indicating that the generated examples cannot decrease the prediction confidence of the target input. 
Larger PCD values mean better test effectiveness.

% In order to further evaluate the probability that the adversarial examples can be reduced, we propose the confidence reduction rate (CRR) as an evaluation metric, as shown in Formula~\ref{eq:crr}:

% \begin{equation}
%   \label{eq:crr}
%   CRR = \mathcal{P}(x)[y] - min \{ \mathcal{P}(x'_1)[y], ..., \mathcal{P}(x'_n)[y] \}
% \end{equation}
% where $x$ is an target input, $x'_i$ is the correspondingly generated adversarial example based on $x$, $y$ is the ground-truth label and $\mathcal{P}( \cdot )$ represents the probability vector generated by the target model.
% In particular, if CRR is less than $0$, we consider that all adversarial examples generated fail to reduce the probability, and CRR will be recorded as $0$. 

Following the existing work~\cite{zhang2022towards, yang2022natural}, we used the \textbf{time spent on the overall testing process} (i.e., completing the testing process for all the subjects) and \textbf{the average number of model invocations for generating examples from a target input}, to measure the efficiency of each technique.
Less time and fewer model invocations mean higher test efficiency.
% Fewer model invocations also have the benefit on security.
% When the model is deployed remotely, frequent model invocations could be identified as malicious attacks and thus result in blocking access to the model~\cite{yang2022natural}.

% \subsubsection{Process}

\subsubsection{Results}
Table~\ref{tab:asr} shows the comparison results among CARROT, ALERT, and \tech{} in terms of RFR.
% \tz{To limit the effects of randomness, we calculated the average results by repeating the experimental procedure ten times.}
% From this table, \tech{} always outperforms CARROT and ALERT on all the tasks based on CodeBERT and GraphCodeBERT, demonstrating the stable attack effectiveness of \tech{}. 
From this table, \tech{} always outperforms CARROT and ALERT on all the subjects, demonstrating the stable effectiveness of \tech{}.
On average, \tech{} improves 70.11\% and 89.83\% higher RFR than CARROT and ALERT across all the five tasks on CodeBERT, 
89.34\% and 57.67\% higher RFR on GraphCodeBERT, 
and 109.26\% and 73.12\% higher RFR on CodeT5, respectively. 

% \tz{On average, \tech{} improves 88.05\% and 72.51\% higher ASR than CARROT and ALERT across all the subjects, respectively.}
% The results show that on the five tasks on CodeBERT, \tech{} is improved by 165.66\%, 114.87\%, 77.88\%, 28.52\% and 32.95\% compared to CARROT, and improved ALERT by 67.06\%, 60.67\%, 120.93\%, 465.14\% and 46.09\%.
% On GraphCodeBERT, \tech{} outperformed CARROT by 153.26\%, 682.00\%, 190.40\%, 34.33\% and 22.13\%, and outperformed ALERT by 23.09\%, 243.84\%, 49.67\%, 411.94\% and 27.30\% on all five tasks.
% \tz{In addition, we hope that the attack technique can not only produce higher ASR, but also generate more diverse adversarial examples. 
% Hence, we analyze diversity in terms of the number of unique successfully-attacking adversarial examples generated by each technique. 
% Larger values for the number of unique successfully-attacking adversarial examples mean better attack effectiveness.}

We then investigated the \textit{unique value} of each technique by analyzing their overlap on target inputs where fault-revealing examples are generated.
On average across all the subjects, there are 30.68\% target inputs where only \tech{} generates fault-revealing examples among the three techniques, while there are just 7.48\% and 2.88\% target inputs where only CARROT and ALERT generate fault-revealing examples, respectively.
The results demonstrate that \tech{} has the largest unique value in revealing faults in deep code models among the three techniques.

We analyzed the effectiveness of \tech{} on \textit{different lengths of code snippets} (ranging from 3 to 8,148 across the five datasets).
We measured the Spearman correlation~\cite{sedgwick2014spearman} between code-snippet length and RFR of \tech{}, and the coefficient is 0.17 (p-value $<$ 0.001). 
That is, there is a weak positive correlation between them.
That indicates the test effectiveness of \tech{} is not significantly affected by code-snippet length, even slightly better on larger code snippets in statistics.
% demonstrating its stable effectiveness.

% \tz{Figure~\ref{fig:unique_successfully_attacking} shows the comparison results among CARROT, ALERT, and \tech{} in terms of the number of unique successfully-attacking adversarial examples on all the subjects. 
% On average, on 27.47\%/8.79\%/2.74\% of target inputs, \tech{}/CARROT/ALERT can attack successfully but any of the other two techniques cannot. 
% From the figure, the average improvements of \tech{} over CARROT and ALERT are 212.51\% and 902.55\% across all the subjects, respectively.
% The results demonstrate that \tech{} has the significantly highest unique value among the three techniques.}

Figures~\ref{fig:pcd_codebert},~\ref{fig:pcd_graphcodebert}, and~\ref{fig:pcd_codet5} show the comparison results among the three techniques in terms of PCD on CodeBERT, GraphCodeBERT, and CodeT5, respectively.
From these figures, the upper quartile, median, and lower quartile of \tech{} are always larger than (or equal to) those of both CARROT and ALERT regardless of the tasks and the pre-trained models, 
demonstrating that the adversarial examples generated by \tech{} can decrease model prediction confidence more significantly. 
% more significant fault-revealing examples for decreasing prediction confidence of target inputs.
For example, on CodeBERT, the average improvements of \tech{} over CARROT and ALERT are 101.88\% and 520.65\% across all the tasks in terms of average PCD, respectively.
% Similarly, on GraphCodeBERT, the average improvements of \tech{} over CARROT and ALERT are 76.35\% and 560.15\%, respectively. 
Similarly, \tech{} improves 76.35\% and 560.15\% higher PCD than CARROT and ALERT on GraphCodeBERT, and 389.10\% and 397.08\% higher PCD on CodeT5, respectively.

% We put the results on all downstream tasks for all models, drawing each box together. 
% As can be seen in Figure~\ref{fig:crr_codebert} and~\ref{fig:crr_graphcodebert}, the CRR value of \tech{} is the largest among all attack techniques, regardless of the model and task. 
% For example, on GraphCodeBERT and Vulnerability Detection, the median CRR of \tech{} is 28.84\%, while that of CARROT and ALERT are 11.64\% and 1.37\%, respectively. 
% It turns out that \tech{} produces more effective adversarial examples.

% Following previous work~\cite{zhang2022towards, yang2022natural}, to evaluate the efficiency of the \tech{}, we choose the time cost, the number of edits, and the invocation number of target model as evaluation metrics.

Besides, CARROT, ALERT, and \tech{} take 290.87 hours, 374.51 hours, and 196.96 hours to complete the entire testing process on all subjects, respectively.
Further, we measured the number of model invocations for each target input during the testing process, whose results are shown in Figures~\ref{fig:inv_codebert},~\ref{fig:inv_graphcodebert}, and~\ref{fig:inv_codet5}. 
% all tasks, respectively, and \tech{} can be implemented in the shortest time possible. 
% \tech{} successfully attacking a code snippet requires an average of 7 edits, while CARROT and ALERT require an average of 8 edits. 
% Finally, we consider the invocation of the target model to estimate the computational cost, which includes forward prediction and backward gradient propagation~\cite{zhang2022towards}. 
% In adversarial attacks, target model invocation can be expensive and time-consuming, and frequent invocations may even be blocked by remotely deployed models, so the number of invocations needs to be kept as low as possible. 
From these figures, \tech{} performs fewer model invocations than both CARROT and ALERT regardless of the tasks and pre-trained models.
On average, \tech{} performs 65.73\% and 78.58\% fewer model invocations than CARROT and ALERT across all the tasks on CodeBERT, 34.07\% and 75.31\% fewer model invocations on GraphCodeBERT, and 52.97\% and 70.09\% fewer model invocations on CodeT5, respectively. 
% \tz{On average, \tech{} performs 55.78\% and 76.45\% fewer model invocations than CARROT and ALERT across all the subjects, respectively.}
The results demonstrate that \tech{} has the significantly highest efficiency among the three techniques.

Overall, the guidance of code differences in \tech{} largely improve test effectiveness and efficiency, which is also the reason why ALERT and CARROT underperform \tech{}. 
Besides, the reason why ALERT usually outperforms CARROT may lie in the former searches more sufficiently, which is confirmed by more model invocations made by ALERT as above.

% \jj{we may use a box to highlight the answer to each RQ.} \tz{done}\jj{seem to have error.}

\begin{tcolorbox}
\textbf{Answer to RQ1:} \tech{} takes less time with fewer model invocations on completing the entire testing process, but generates more fault-revealing examples with more significant prediction confidence decrement on all the subjects, than the state-of-the-art baselines.
\end{tcolorbox}

\begin{table*}[]
    \caption{Robustness enhancement of the target models after adversarial fine-tuning}
    % \vspace{-2mm}
    \label{tab:retraining}
    \centering
    \tabcolsep=1.0mm
    \begin{adjustbox}{max width=1.0\textwidth,center}
        \begin{tabular}{ cc|ccc|ccc|ccc|ccc }
            \toprule
        	\multirow{2}{*}{\textbf{Task}} & \multirow{2}{*}{\textbf{Model}} & \multicolumn{3}{c|}{\textbf{Ori}} & \multicolumn{3}{c|}{\textbf{CARROT}} & \multicolumn{3}{c|}{\textbf{ALERT}} & \multicolumn{3}{c}{\textbf{\tech{}}} \\
        	\cmidrule(lr){3-5} \cmidrule(lr){6-8} \cmidrule(lr){9-11} \cmidrule(lr){12-14}
        	& & CARROT & ALERT & \tech{} & CARROT & ALERT & \tech{} & CARROT & ALERT & \tech{} & CARROT & ALERT & \tech{} \\
        	\midrule
        	\multirow{3}{*}{\makecell[c]{Vulnerability \\ Prediction}} & CodeBERT & 62.96\% & 62.77\% & \textbf{63.03\%} & 29.14\% & 21.11\% & \textbf{29.69\%} & 23.43\% & 26.27\% & \textbf{34.44\%} & 32.16\% & 31.73\% & \textbf{38.82\%} \\
             & GraphCodeBERT & \textbf{62.99\%} & 62.88\% & 62.92\% & 12.37\% & 19.59\% & \textbf{21.65\%} & 16.33\% & 17.35\% & \textbf{23.71\%} & 25.77\% & 24.74\% & \textbf{34.02\%} \\
             & CodeT5 & 63.69\% & 63.81\% & \textbf{63.92\%} & 52.03\% & 39.76\% & \textbf{82.03\%} & 42.26\% & \textbf{49.11\%} & 44.26\% & 41.43\% & 45.52\% & \textbf{52.54\%} \\ \hline
        	\multirow{3}{*}{\makecell[c]{Clone \\ Detection}} & CodeBERT & 97.39\% & 96.45\% & \textbf{97.45\%} & 83.15\% & 42.31\% & \textbf{94.44\%} & 52.65\% & 72.46\% & \textbf{75.32\%} & 38.51\% & 71.45\% & \textbf{89.78\%} \\
        	 & GraphCodeBERT & 97.01\% & 97.22\% & \textbf{97.43\%} & 75.00\% & 66.67\% & \textbf{77.50\%} & 79.17\% & 84.29\% & \textbf{92.31\%} & 35.71\% & 57.69\% & \textbf{92.97\%} \\
            & CodeT5 & 97.73\% & 97.14\% & \textbf{98.10\%} & 67.77\% & 57.63\% & \textbf{75.85\%} & 69.94\% & 64.36\% & \textbf{81.63\%} & 42.15\% & 51.74\% & \textbf{79.88\%} \\ \hline
        	\multirow{3}{*}{\makecell[c]{Authorship \\ Attribution}} & CodeBERT & 90.55\% & 89.39\% & \textbf{90.91\%} & \textbf{45.06\%} & 40.67\% & 41.03\% & 51.25\% & 56.25\% & \textbf{58.82\%} & 45.67\% & 43.33\% & \textbf{76.47\%} \\
        	 & GraphCodeBERT & 89.39\% & 88.72\% & \textbf{90.35\%} & \textbf{81.75\%} & 67.08\% & 72.40\% & 79.41\% & 78.67\% & \textbf{100.00\%} & 45.59\% & 80.39\% & \textbf{84.75\%} \\
            & CodeT5 & 92.43\% & 92.68\% & \textbf{93.03\%} & 70.95\% & 65.91\% & \textbf{73.48\%} & 55.73\% & 71.88\% & \textbf{76.44\%} & 44.31\% & 52.56\% & \textbf{72.37\%} \\ \hline
        	\multirow{3}{*}{\makecell[c]{Functionality \\ Classification}} & CodeBERT & 98.11\% & 98.52\% & \textbf{98.56\%} & \textbf{83.46\%} & 72.80\% & 81.51\% & 70.83\% & 71.75\% & \textbf{79.41\%} & 78.92\% & 71.18\% & \textbf{95.43\%} \\
        	 & GraphCodeBERT & 98.48\% & 98.55\% & \textbf{98.72\%} & 67.53\% & 75.19\% & \textbf{77.27\%} & 32.04\% & 52.62\% & \textbf{62.98\%} & 91.22\% & 90.81\% & \textbf{93.08\%} \\
            & CodeT5 & 97.92\% & 98.46\% & \textbf{98.63\%} & 25.31\% & 21.33\% & \textbf{27.36\%} & 41.07\% & 57.14\% & \textbf{57.42\%} & 24.87\% & 59.58\% & \textbf{63.76\%} \\ \hline
        	\multirow{3}{*}{\makecell[c]{Defect \\ Prediction}} & CodeBERT & 83.50\% & 84.16\% & \textbf{84.44\%} & 52.73\% & 25.81\% & \textbf{66.03\%} & 74.88\% & 75.87\% & \textbf{83.12\%} & 76.86\% & 68.66\% & \textbf{85.36\%} \\
        	 & GraphCodeBERT & 83.34\% & 84.00\% & \textbf{84.53\%} & 68.20\% & 48.54\% & \textbf{74.88\%} & 52.73\% & \textbf{63.91\%} & 59.45\% & 67.08\% & 68.66\% & \textbf{76.14\%} \\ 
            & CodeT5 & 80.92\% & 81.32\% & \textbf{81.57\%} & 31.48\% & 34.08\% & \textbf{37.73\%} & 31.75\% & 42.22\% & \textbf{55.77\%} & 54.45\% & 54.18\% & \textbf{73.83\%} \\ \midrule
        	\multicolumn{2}{c|}{Average} & 86.43\% & 86.40\% & \textbf{86.91\%} & 56.40\% & 46.57\% & \textbf{62.19\%} & 51.56\% & 58.94\% & \textbf{65.67\%} & 49.65\% & 58.15\% & \textbf{73.95\%} \\
            \bottomrule
        \end{tabular}
    \end{adjustbox}
    % \vspace{-6mm}
\end{table*}

% \vspace{-1mm}
\subsection{RQ2: Model Robustness Enhancement}
\label{sec:rq2}
% \vspace{-0.5mm}
\subsubsection{Setup}
We studied the value of generated adversarial examples by using them to enhance the robustness of the target model via an adversarial fine-tuning strategy.
For each subject, we divided the test set into two equal parts ($S_1$ and $S_2$), to avoid data leakage between the augmented training set and the evaluation set constructed by the same technique.
Specifically, we applied each technique to generate examples from $S_1$, and obtained an fault-revealing example or an example that produces the largest decrement on prediction confidence (if no fault-revealing example is generated) for each target input.
These examples were integrated with the training set to form the augmented training set, which is used for fine-tuning the model.
Hence, for a given subject, the size of the augmented training set constructed by each technique is the same.

After obtaining a fine-tuned model for each subject with each technique, we evaluated it on the evaluation set of the fault-revealing examples generated from $S_2$ by \tech{}, CARROT, and ALERT, respectively.
% Then, we measured the accuracy of the fine-tuned model on the three evaluation sets to measure its ability of defending against attacks.
Then, we measured the accuracy of the fine-tuned model on the three evaluation sets to measure its ability of reducing faults.
\subsubsection{Results}
Table~\ref{tab:retraining} shows the effectiveness of enhancing model robustness with the generated examples by the studied techniques, respectively.
The first row (except Column Ori) represents the evaluation set constructed by the corresponding technique, while the second row represents the augmented training set constructed by the corresponding technique.
Column Ori lists the accuracy of the fine-tuned model on the original test set.
% the fine-tuned model is \textit{evaluated} on the successfully-attacking adversarial examples generated by the corresponding technique (denoted as $T_e$),
% and the second row represents the fine-tuned model is \textit{obtained} based on the examples generated by the corresponding technique (denoted as $T_f$) \tz{not clear}.
%\tz:1
The values in the columns (except Column Ori) represent the ratio of the faults (revealed by the evaluation set) that can be eliminated by the fine-tuned model based on the augmented training set.
% We found on most subjects, \tech{} improves the model robustness to defend against attacks from the largest ratio of adversarial examples generated by \tech{}, CARROT, ALERT, respectively.
We found that on most subjects, \tech{} enhance the model robustness to reduce the largest ratio of faults revealed by \tech{}, CARROT, ALERT, respectively.
% On average, the models fine-tuned by \tech{} can defend against attacks from 62.19\%, 65.67\%, 73.95\% of successfully-attacking examples generated by CARROT, ALERT, \tech{} respectively, with the improvement of 10.27\%, 27.37\%, 48.94\% over those by CARROT and 33.54\%, 11.42\%, 27.17\% over those by ALERT respectively.
On average, the models fine-tuned by \tech{} can reduce 62.19\%, 65.67\%, 73.95\% of faults revealed by CARROT, ALERT, and \tech{} respectively, with the improvement of 10.27\%, 27.37\%, 48.94\% over those by CARROT and 33.54\%, 11.42\%, 27.17\% over those by ALERT respectively.
% Besides, the results of attack defense between different techniques indicate that the examples generated by \tech{} could subsume those by CARROT and ALERT to a large extent.
Besides, the results of robustness enhancement between different techniques indicate that the examples generated by \tech{} could subsume those by CARROT and ALERT to a large extent.
In five cases, \tech{} performs worse than ALERT or CARROT, as the augmented training set and the evaluation set generated by the same technique could share a higher degree of similarity (facilitating fine-tuning).
% \tz{
% In five cases, \tech{} underperforms compared to ALERT or CARROT due to potential overfitting caused by augmented training sets and evaluation sets that share a higher degree of similarity within the same technique.}\tz{It need more explanation why CODA performs worse than baselines}

By comparing Column Ori in Table~\ref{tab:retraining} and the last column in Table~\ref{tab:tasks_and_models}, the original model and the fine-tuned model via \tech{} have close accuracy, i.e., all the absolute accuracy differences are less than 1\%.
The results demonstrate \tech{} is more helpful to improve model robustness than CARROT and ALERT without damaging the original model performance.
% \tz{Do we need to explain bad cases?}

% the latter may provide more similar adversarial training data to the corresponding evaluation set.

\begin{tcolorbox}
\textbf{Answer to RQ2}: \tech{} helps enhance the model robustness more effectively than CARROT and ALERT, in terms of reducing faults revealed by the examples generated by itself as well as the examples generated by the other two techniques.
\end{tcolorbox}

% Notably, the original model performed well on the original testing set, but mispredicted all adversarial examples on the other three adversarial testing sets. 
% Table~\ref{tab:retraining} shows the specific prediction accuracy. 
% We can observe that all the retrained models perform close to the original models on the original dataset with an error of less than 1.00\%, indicating that the retraining process does not lose much of the original performance.
% All adversarial fine-tuned models perform much better than the original models on the three adversarial testing sets. 
% The average accuracies of CARROT on the three adversarial testing sets are 59.84\%, 53.27\% and 53.75\%, and ALERT are 47.98\%, 59.94\% and 60.86\%, respectively. 
% \tech{} has the best overall performance with 63.64\%, 66.96\% and 76.68\% respectively.

% \vspace{-1mm}
\subsection{RQ3: Contribution of Main Components}
\label{sec:rq3}

\subsubsection{Setup}
We studied the contribution of each main component in \tech{}, i.e., reference inputs selection (RIS), equivalent structure transformations (EST), and identifier renaming transformations (IRT).
We constructed four variants of \tech{}:
\begin{itemize}
    \item \textbf{w/o RIS}: we replaced RIS with the method that randomly selects $N$ inputs from training data as reference inputs.
    
    \item \textbf{w/o EST}: we removed EST from \tech{}, i.e., it directly performs identifier renaming transformations after selecting reference inputs.

    \item \textbf{w/o CDG} (code difference guidance in EST): we replaced the code-difference-guided strategy used for EST in \tech{} with randomly selecting rules for EST.
    
    % with the method that randomly selects structure transformations from the collection of potential transformation rules.}
    
    \item \textbf{w/o IRT}: we removed IRT from \tech{}, i.e., it directly checks whether a fault-revealing example is generated after equivalent structure transformations.
\end{itemize}

% To explore the contribution of individual components in \tech{}, we performed ablation experiments on the \tech{}, removing each component separately to determine the contribution of each component. 

\subsubsection{Results}
Table~\ref{tab:ablation} shows the average RFR values of each technique across all the tasks on CodeBERT, GraphCodeBERT, and CodeT5, respectively.
The results on each task can be found at our project homepage~\cite{coda2022} due to the space limit.
\tech{} outperforms all four variants in terms of average RFR with improvements of 15.79\%$\sim$165.27\%,
% \tz{absolute values: 10.76\%$\sim$43.17\%.}
demonstrating the contribution of each main component in \tech{}.
Also, reference inputs selection and identifier renaming transformations contribute more than equivalent structure transformations.
The possible reason is that not all the rules of equivalent structure transformations can be applicable to all the target inputs, but identifier renaming transformations are applicable to all the inputs.
We can enrich the rules of equivalent structure transformations in the future to further improve the test effectiveness.
The comparison results among \tech{}, w/o EST, and w/o CDG demonstrate the contribution of our code-difference-guided strategy for applying equivalent structure transformations for testing deep code models.
Besides, ALERT targets only identifier-level adversarial example generation, and thus we further compared ALERT with w/o EST for fairer comparison.
The results also demonstrate the superiority of the latter, showing the effectiveness of our code difference guided adversarial example generation (despite only considering identifier renaming transformation).

% e also found that w/o EST achieves higher RFR than ALERT on average across all the five datasets, demonstrating the effectiveness of \tech{} for identifier-level adversarial example generation.}

% \tz{

% 1. Possibly less codes can apply structure transformation, but almost all codes can apply identifier renaming.

% 2. An adv code may only change the structure 1 time, but change the variable name 7 times on average.

% 3. The number of structural transformation rules we currently design is limited, however the number of variable renaming options is much larger than the number of structural transformation rules.
% }

\begin{tcolorbox}
\textbf{Answer to RQ3}: All the components of reference input selection, equivalent structure transformations, and identifier renaming transformations make contributions to the overall effectiveness of \tech{}, demonstrating the necessity of each of them in \tech{}.
\end{tcolorbox}

\begin{table}[]
    \caption{Ablation test for \tech{} in terms of average RFR}
    % \vspace{-2mm}
    \label{tab:ablation}
    \centering
    \tabcolsep=1.2mm
    \begin{adjustbox}{max width=0.47\textwidth,center}
        \begin{tabular}{ cccccc }
            \toprule
        	\textbf{Model} & \textbf{w/o RIS} & \textbf{w/o EST} & \textbf{w/o CDG} & \textbf{w/o IRT} & \textbf{\tech{}} \\
        	\midrule
        	 CodeBERT & 30.83\% & 62.73\% & 63.08\% & 35.14\% & \textbf{73.04\%} \\
        	GraphCodeBERT & 29.49\% & 62.41\% & 61.98\% & 26.24\% & \textbf{73.62\%} \\
            CodeT5 & 26.75\% & 50.74\% & 57.98\% & 38.21\% & \textbf{70.96\%} \\
            \bottomrule
        \end{tabular}
    \end{adjustbox}
    % \vspace{-2mm}
\end{table}
% \vspace{-0.5mm}
\subsection{RQ4: Naturalness of Adversarial Examples}
\label{sec:rq4}

\subsubsection{Setup}
It is important to check whether the generated fault-revealing examples are natural to human judges~\cite{yang2022natural,szegedy2014intriguing}.
Here, we conducted a user study to compare the naturalness of examples generated by \tech{}, CARROT, and ALERT, and \textit{our user study shares the same design as the one conducted by the existing work~\cite{yang2022natural}}:

\textbf{Data Preparation.} For each subject, we randomly sampled 10 target inputs, and then for each technique on each target input, we randomly sampled a generated example.
% from the set of examples generated from each sampled target input.
% \jj{plz check: For each subject, we randomly sampled 10 target inputs, and then for each technique we randomly sampled an adversarial example from the set of examples generated from each sampled target input.}
That is, for each sampled target input, we construct three pairs of code snippets, each of which contains the target input and an adversarial example generated by \tech{}, CARROT, or ALERT.
In total, we obtained 450 pairs of code snippets for the user study due to 15 subjects $\times$ 3 techniques.

\textbf{Participants.} Same as the existing work~\cite{yang2022natural}, the user study also involves four non-author participants, each of whom has a Bachelor/Master degree in Computer Science with at least five years of programming experience. 

\textbf{Process.} For objective evaluation, we did not tell participants which technique generates the adversarial example in a pair of code snippets.
Also, we highlighted the changes in each pair of code snippets for facilitating manual evaluation.
Then, each participant individually evaluated each pair by evaluating to what extent the changes are natural to the code context and the changed identifiers preserve the original semantics, following the existing work~\cite{yang2022natural}.
Specifically, participants gave a score for each pair based on a 5-point Likert scale~\cite{joshi2015likert} (1 means strongly disagree and 5 means strongly agree).
% following the existing work~\cite{jin2020bert, yang2022natural}.
More details about the design can be found in the existing work~\cite{yang2022natural}.

\begin{figure}[t]
    \centering
    \includegraphics[width=1.0\linewidth]{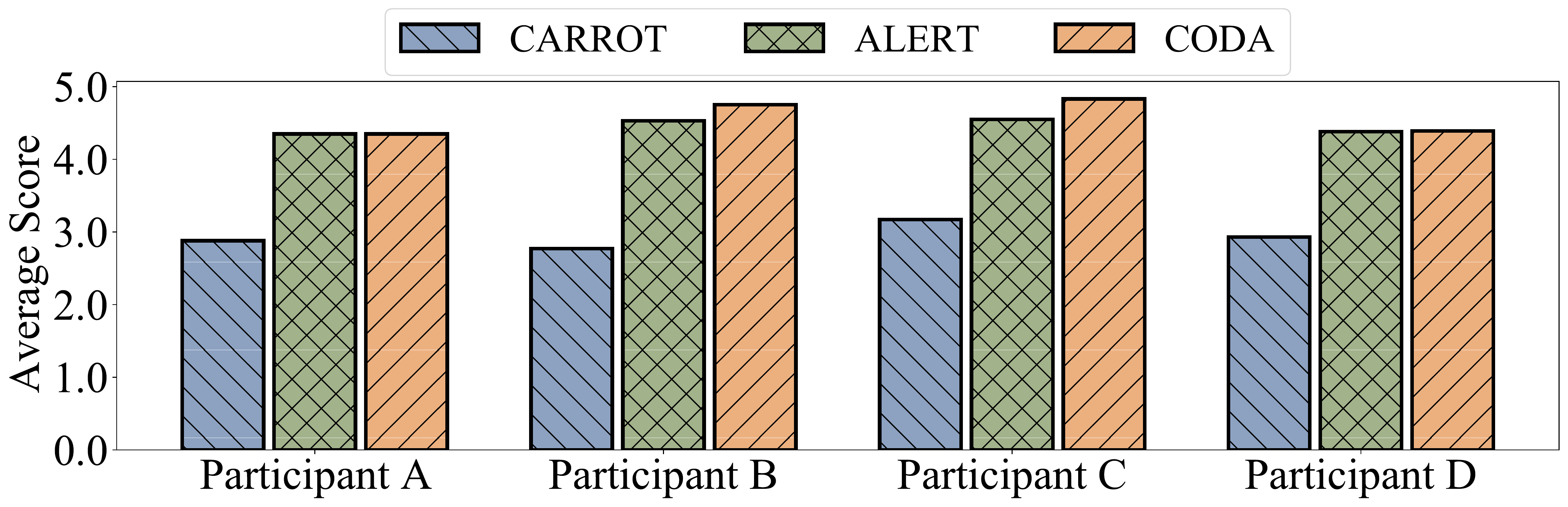}
    % \vspace{-4mm}
    \caption{ Average score to 
    evaluate naturalness of examples}
    \label{fig:naturalness}
    % \vspace{-2mm}
\end{figure}

% , where the x-axis represents each participant and the y-axis shows the average score. 
% The results show that ALERT and \tech{} can generate more natural adversarial examples. 
% The average rating of the four participants on the adversarial examples generated by ALERT and \tech{} was close to 4.50, indicating that the participants believed that the replacements produced by these two methods were natural. 
% Participants consistently gave the adversarial examples generated by CARROT low scores (2.89 on average), indicating that they thought the adversarial examples of CARROT were unnatural.

\subsubsection{Results}
Figure~\ref{fig:naturalness} shows the average score of the examples generated by each technique for each participant.
The conclusions from different participants are consistent: the naturalness of the examples generated by \tech{} and ALERT is closely high (round 4.50 on average), and significantly higher than that by CARROT (just 2.94 on average).
% \tz{Moreover, the p-value of CARROT and \tech{} is 0.02 (that is less than 0.05), while the p-value of ALERT and \tech{} is 0.63 (that is greater than 0.05), and thus the results demonstrate that \tech{} has a statistically significant superiority over CARROT.}
ALERT is a naturalness-aware technique, whose core contribution is to ensure naturalness of generated examples, but \tech{} achieves similar naturalness scores to it, demonstrating that \tech{} can generate highly natural adversarial examples.

\begin{tcolorbox}
\textbf{Answer to RQ4}: The adversarial examples generated by \tech{} are natural closely to the state-of-the-art naturalness-aware adversarial example generation technique (i.e., ALERT), which is consistently confirmed by participants.
\end{tcolorbox}
% \vspace{-1mm}
\section{Threats to Validity}
\label{sec:dis}

\begin{table}[]
    \caption{Influence of hyper-parameter $U$.}
    % \vspace{-2mm}
    \label{tab:n_1}
    \centering
    \tabcolsep=2.0mm
    \begin{adjustbox}{max width=0.47\textwidth,center}
        \begin{tabular}{ cccccc }
            \toprule
        	\textbf{U} & \textbf{64} & \textbf{128} & \textbf{256} & \textbf{512} & \textbf{1024} \\
        	\midrule
        	CodeBERT & 60.14\% & 67.90\% & 73.04\% & 75.27\% & 75.83\%  \\ 
            GraphCodeBERT & 61.92\% & 70.16\% & 73.62\% & 74.98\% & 75.69\%  \\ 
            CodeT5 & 51.22\% & 61.74\% & 70.96\% & 72.96\% & 76.88\%  \\ 
            \bottomrule
        \end{tabular}
    \end{adjustbox}
    % \vspace{-5mm}
\end{table}

\begin{table}[t]
    \caption{Influence of hyper-parameter $N$}
    % \vspace{-2mm}
    \label{tab:n_2}
    \centering
    \begin{adjustbox}{max width=0.49\textwidth,center}
        \begin{tabular}{ cccccccc }
            \toprule
        	\textbf{N} & \textbf{1} & \textbf{4} & \textbf{16} & \textbf{32} & \textbf{64} & \textbf{128} \\
        	\midrule
            CodeBERT & 28.08\% & 46.33\% & 61.07\% & 67.12\% & 73.04\% & 76.38\% \\ 
            GraphCodeBERT & 31.84\% & 46.46\% & 60.40\% & 66.12\% & 73.62\% & 74.93\% \\ 
            CodeT5 & 27.05\% & 43.71\% & 58.19\% & 64.82\% & 70.96\% & 73.25\% \\ 
            \bottomrule
        \end{tabular}
    \end{adjustbox}
    % \vspace{-5mm}
\end{table}

% \subsection{The influence of hyper-parameters}

The main threat to validity lies in the settings of parameters in \tech{}.
Here, we investigated the influence of two important parameters in \tech{} (i.e., $U$ and $N$ introduced in Section~\ref{sec:ReferenceInputsSelection}).
They affect the selection of reference inputs.
% We mentioned two important hyper-parameters $N_1$ and $N_2$ in Section~\ref{sec:ReferenceInputsSelection}.
% These two hyper-parameters affect the quality and quantity of the reference inputs, which in turn affects the quality of the adversarial examples generated by \tech{}.
Tables~\ref{tab:n_1} and~\ref{tab:n_2} show the influence of $U$ and $N$ in terms of average RFR across all the tasks.
As $U$ increases, \tech{} performs better, as incorporating more inputs for the selection based on similarity can increase the possibility of finding more effective reference inputs.
Similarly, as $N$ increases within our studied range, more effective ingredients could be included, leading to better effectiveness.
However, the amount of increase in terms of average RFR becomes smaller with $U$ and $N$ increasing, and meanwhile incorporating more inputs can incur more costs in similarity calculation or code transformations.
Hence, by balancing the effectiveness and efficiency of \tech{}, we set $U$ to 256 and $N$ to 64 as the default settings in \tech{} for practical use.

\section{Related Work}
\label{sec:related}
% \tz{explain why we did not select these baselines?}
Besides the state-of-the-art techniques compared in our study (i.e., CARROT~\cite{zhang2022towards} and ALERT~\cite{yang2022natural}), there are some other adversarial example generation techniques for deep code models~\cite{chen2022generating,srikant2021generating,zhang2020generating}.
For example, Yefet et al.~\cite{yefet2020adversarial} proposed DAMP, which changes variables in the target input by gradient computation.
It only works for the models using one-hot encoding to process code, and thus cannot generate adversarial examples for the models based on state-of-the-art CodeBERT~\cite{feng2020codebert}, GraphCodeBERT~\cite{guo2021graphcodebert}, and CodeT5~\cite{wang2021codet5} due to different encoding methods.
% , not to deep code models that use BPE~\cite{gage1994new, liu2019roberta} to process tokens.
Zhang et al.~\cite{zhang2020generating} proposed MHM, which iteratively performs identifier renaming transformations to generate adversarial examples based on the Metropolis-Hastings~\cite{metropolis1953equation, hastings1970monte, chib1995understanding} algorithm.
MHM underperforms CARROT and ALERT as presented by the existing studies~\cite{zhang2022towards,yang2022natural}.
% However, MHM cannot generate natural disturbances and is computationally expensive.
Pour et al.~\cite{pour2021search} proposed a search-based technique with an iterative refactoring-based process.
% based on code refactoring and mutation testing guidance \jj{not clear}. 
% \tz{Pour et al.~\cite{pour2021search} proposed a search-based testing technique for code embedding models to evaluate their robustness. This technique uses an iterative guided refactoring process to generate adversarial examples.}
% This algorithm still does not take into account the naturalness of the generated adversarial examples, and dead code insertion is completely unnatural.
% In these studies, two algorithms that have been shown to be more efficient than others are CARROT~\cite{zhang2022towards} and ALERT~\cite{yang2022natural}, which we use as our baselines.
% We have introduced the details of these two state-of-the-art algorithms in Section ~\ref{sec:techniques}.
% \tz{
% Chen et al.~\cite{chen2022generating} apply 6 simple predefined code transformation rules on Java to attack code models.
% However, they just iterate over all available transforms to choose the best one without any optimization.
Ramakrishnan et al.~\cite{henke2022semantic} generated adversarial examples for deep code models via gradient-based optimization, including renaming transformations and dead code insertion.
% It also does not ensure the naturalness of generated examples, especially without the real variable names.
% Ramakrishnan et al.~\cite{henke2022semantic} attack code models by gradient-based optimization of the renaming transformation, but the rule of dead code insertion does not ensure the naturalness of generated examples.
% Srikant et al.~\cite{srikant2021generating} uses optimized program obfuscations (e.g., identifier renaming and dead code insertion) to modify the code, which also does not ensure the naturalness of generated examples.
% }
% Jha et al.~\cite{jha2022codeattack} adopted the masked programming language model with greedy search to predict substitutes for vulnerable tokens.
Jha et al.~\cite{jha2022codeattack} proposed CodeAttack, which adopts the masked language model with greedy search to predict substitutes for vulnerable tokens.
All of them do not ensure the naturalness of generated examples, especially with the rule of dead code insertion.
% Similarly, they also do not ensure the naturalness of generated examples.
Also, these techniques still search for ingredients in the enormous space, limiting their effectiveness.
Different from them, our work designs the first code-difference-guided adversarial example generation technique, which can largely reduce ingredient space for improving the test effectiveness.
\section{Conclusion and Future Work}
To improve test effectiveness on deep code models, we propose a novel perspective by exploiting the code differences between reference inputs and the target input to guide the generation of adversarial examples.
From this perspective, we design \tech{}, which reduces the ingredient space as the one constituted by structure and identifier differences and designs equivalent structure transformations and identifier renaming transformations to preserve original semantics.
% We introduce \tech{}, which exploits the code differences between reference inputs and target input to guide the adversarial attack process of deep code models. 
% This technique can greatly reduce the ingredient space, thus improving the attack efficiency. 
We conducted an extensive study on 15 subjects.
The results demonstrate that \tech{} reveals more faults with less time than the state-of-the-art techniques (i.e., CARROT and ALERT), and confirm the capability of enhancing the model robustness.
% Experimental results show that \tech{} can effectively generate adversarial examples to attack state-of-the-art models with high attack success rate. 
% Moreover, \tech{} achieves an efficient attack because it has the less running time and the fewer model invocations. 
% A user study confirmed that \tech{} can generate adversarial examples that look natural to human judges. 
% Finally, we explore the effect of adversarial examples generated by \tech{} on enhancing the robustness of deep code models via an adversarial fine-tuning strategy.
% In the future, we will expand and evaluate more deep code models and more tasks for further research.

In the future, we can improve \tech{} from several aspects.
First, \tech{} successively applies equivalent structure transformations and identifier renaming transformations without backtracking.
In the future, we can improve it by backtracking and trying more rounds of transformations if the first round of equivalent structure transformations and identifier renaming transformations fails, which may help transform more target inputs to fault-revealing examples.
Second, the number of inputs belonging to the class with the second largest probability may be not enough, even though we did not come across this case in our evaluation.
If it occurs in the future, \tech{} may use the next highest-probability classes as a compromise.
Third, we will explore the use of test data to provide reference inputs to further improve \tech{}.
Fourth, we will investigate some other strategies (such as the type obfuscation strategy~\cite{compton2020embedding}) to replace the simple strategy using the placeholder {\tt <unk>} in RIS, in order to further improve the effectiveness of \tech{}.

% our focus will be on enhancing the code similarity strategy for RIS modules. 
% Compton et al.~\cite{compton2020embedding} proposed the type obfuscation strategy during model training,
% which enforces the reliance of model on code structure over specific identifiers. 
% In the future, we aim to investigate the synergies between the type obfuscation strategy and \tech{}.}

\section*{Acknowledgements}
This work was supported by the National Natural Science Foundation of China Grant Nos. 62322208, 62002256, 62192731, 62192730, and CCF Young Elite Scientists Sponsorship Program (by CAST),

% This work is partially supported by the National Natural Science Foundation of China under Grant Nos. 62192731 and 62192730.\tz{check and update, plz}

\balance
\bibliographystyle{IEEEtrans}
\bibliography{reference}

\end{document}